\documentclass[a4paper,11pt]{article}
\pdfoutput=1 

\usepackage{jcappub} 

\usepackage[T1]{fontenc} 

\usepackage{natbib}
\usepackage{graphicx}
\usepackage{caption}
\usepackage{tensor}
\usepackage{subcaption}
\usepackage{enumerate}
\usepackage{dsfont}

\title{\boldmath Effects of Reheating on Moduli Stabilization }

\author[a]{Khursid Alam,}
\author[a]{and Koushik Dutta}

\affiliation[a]{Indian Institute of Science Education and Research Kolkata, Mohanpur, WB741246, India}
\emailAdd{khursid@iiserkol.ac.in}
\emailAdd{koushik@iiserkol.ac.in}

\abstract{Moduli potential loses its minima due to external energy sources of inflaton energy density or radiation produced at the end of inflation. But, the non-existence of minima does not necessarily mean destabilization of moduli. In fact, the destabilization of moduli is always dependent on the initial field values of the fields. In this work, we study carefully how the effects of reheating ease the problem of moduli destabilization. The associated time scale to produce the thermal bath allows a larger initial field range to stabilize the field. Contrary to the usual notion, the allowed initial field range is larger for higher temperatures when the effective potential is of a run-away nature. This eases the moduli destabilization problem for heavy mass moduli. For low mass moduli ($\lesssim$ 30 TeV), the allowed field range still causes the cosmological moduli problem by violating the BBN constraints unless its initial abundance is suppressed.}

\begin{document}
\maketitle
\flushbottom


\section{Introduction}
\label{sec:intro}

In supersymmetric theories beyond the Standard Model, there exist several massless scalar fields. To avoid stringent fifth-force constraints, these fields must be massive, and their mass is usually related to the effects of supersymmetry breaking. In the context of supergravity, these fields are gravitationally coupled with other fields whose decay widths are Planck suppressed. For our considerations, we will call these fields collectively `moduli' represented by $\sigma$ (or its canonically normalized version $\phi$). In the context of String Theory, these fields characterize either the value of the low energy gauge/Yukawa coupling constants or the volume of the compact internal manifolds. Due to phenomenological constraints related to the time variations of the coupling constants or to avoid the decompactification of internal spaces, it is crucial that the vev of these fields are fixed at finite values. This requires a clear understanding of the sources of the potential in some fundamental theory, as well as the evolution of these fields in a cosmological background \cite{Binetruy:2006ad}.

The typical decay widths of these moduli fields are given by $\Gamma_{\phi} \sim m_\phi ^3/M_{\rm pl}^2$, where $m_\phi$ is the mass of these fields. Therefore, if the field is lighter than $20$ MeV, its decay time is larger than the age of the Universe. On the other hand, unless the field is lighter than $10^{-26}$ eV with an initial amplitude of $\sim M_{\rm pl}$, it carries too much energy to overclose the Universe. These scalar fields typically remain away from their global minimum and when their masses become the order of the Hubble constant, they start to oscillate \cite{Dine:1995uk}, \cite{Cicoli:2016olq}. The energy densities carried by these fields redshift slower than any preexisting radiation, and soon may start to dominate the energy density of the Universe. But, as long as the fields are heavier than $\sim 30$ TeV, it decays well before the Big Bang Nucleosynthesis (BBN), and it is cosmologically consistent with all observations. In several models in supergravity and String Theory, moduli have masses in the range of few GeV to TeV, and it is related to the supersymmetry breaking scale. In this case, unless moduli abundance is highly suppressed i.e $n_\phi/ s \lesssim 10^{-12}$, the moduli will decay during or after BBN whose decay products spoil the light element abundance. Here, $n_{\phi}$ is moduli number density and $s$ is the entropy density. In the literature, the problems created by moduli with masses of few GeV to TeV is dubbed as cosmological moduli problem \cite{Coughlan:1983ci}, \cite{Banks:1993en}, \cite{deCarlos:1993wie}.

As mentioned before, the moduli fields at their present minimum must be massive. Moreover, the value of the potential at the minimum needs to be the present value of the cosmological constant required to explain the recent cosmic acceleration. In several String Theory constructions, the potential also has another minimum at an infinite field value, and the two minima are separated by a finite barrier height \cite{Kachru:2003aw}. In some cases, additionally, there might be another AdS global minimum, and in that case, the typical tunneling time must be smaller than the age of the Universe \cite{Kallosh:2004yh}. A schematic moduli potential (blue line) is shown in Fig.~\ref{fig:KKLT_potential} where the de-Sitter minimum at a finite field value is at $\sigma_{\rm min}$, and the barrier is at $\sigma_{\rm max}$ with potential value $V_{\rm max}$. In the plot, we have shown the potential in terms of the canonically normalized field $\phi$.

Even if the modulus potential has desired properties, it is important that the modulus field remains stabilized at a finite field value at $\sigma_{\rm min}$. This is not necessarily guaranteed in a cosmological setup. For example, even if the potential is of the above form with the required minimum at a finite field value, the field may start to evolve from high up the potential on the left-hand side of the minimum where the potential is very steep. In this case, even with cosmological damping, the field might have enough kinetic energy to overshoot the finite barrier height \cite{Brustein:1992nk}. Even worse, if the field has an initial field value greater than $\sigma_{\rm max}$, the field is guaranteed to eventually reach the destabilization limit of the infinite value.  The issue of overshooting the barrier due to initial field configuration is called the `Brustein and Steinhardt' problem in the literature  \cite{Brustein:1992nk}. Several solutions to the problem have been proposed in the literatures \cite{Barreiro:1998aj}, \cite{Brustein:2004jp}, \cite{Kaloper:2004yj} \cite{Barreiro:2005ua}, and all are related to the idea of adding additional background energy densities. Therefore, dynamical analysis with the sensitivity of the initial conditions is necessary to understand the issue of moduli stabilization.  

Let us call the moduli potential with the above-mentioned properties $V_0 (\sigma)$. But, in any realistic theory, this potential is not in isolation. In fact, in the context of inflation in supergravity and in String Theory, the moduli fields are generically coupled with the inflaton $\varphi$. The nature of the coupling is dictated by the supergravity, and the total potential during inflation in the simplest form roughly looks like

\begin{equation} \label{v_total_inf}
V_{\rm total} = V_0 (\sigma) + V_{\rm inf} (\varphi, \sigma)~,
\end{equation}

where the coupling term typically is of the form $V_{\rm inf} (\varphi, \sigma) = V(\varphi)/ \sigma^n$ with $ n > 0$. Here $V(\varphi)$ drives inflation. For any explicit model, the potential for $V_{\rm total}$ can be much more complicated than the separable form of Eq.~\eqref{v_total_inf}. But, the crucial point is that if $V(\varphi)$ is large enough, $V_{\rm total}$ will have a run-away direction along the modulus field  \cite{Kallosh:2004yh}. In particular, when $V(\varphi) \sim V_{\rm max}$, the potential completely loses its minimum at a finite field value. In this case, it was argued that the modulus field will eventually roll towards the large vev. This in fact leads to the famous KL bound of the Hubble scale during inflation $H_{\inf} \lesssim m_{3/2}$ for the KKLT model  \cite{Kallosh:2004yh}. It is important to note that the bound arises due to the finite barrier height and coupling between the inflation and the moduli sector.

There is another source of destabilization of the modulus potential, and that is the main topic of this work. At the end of inflation, the energy density stored in the inflaton decays to some lighter species producing a hot thermal plasma in the end. The thermal plasma induces a temperature-dependent potential for the modulus field of the form $V_{\rm thermal} \sim T^4/ \sigma$ where

\begin{equation} \label{v_total_thermal}
V_{\rm total} = V_0 (\sigma) + V_{\rm thermal} (T, \sigma)~.
\end{equation}

As like the inflaton dependent term above, it is clear that for high enough temperature $T_{\rm crit} \sim V_{\rm max}^{1/4} $, the minima will disappear and the full potential will have a run-away behavior \cite{Buchmuller:2004xr}, \cite{Buchmuller:2004tz}. In fact, this is the original argument of having a maximum reheating temperature which spoils the moduli stabilization structure of the zero temperature potential of $V_0 (\sigma)$. In this case, it was assumed that as soon as the potential loses its local minimum, the field will destabilize leading to infinite vev in the far future. The thermal corrections to the moduli potential always induce some initial misalignment of the field, and it leads to a certain amount of the modulus coherent oscillations \cite{Nakayama:2008ks}. For lighter mass modulus this resurrects the usual cosmological moduli problem. In the context of LARGE Volume type IIB flux compactifications, the finite-temperature corrections to the modulus potential have been calculated explicitly in \cite{Anguelova:2009ht} and some cosmological implications have been discussed in \cite{Anguelova:2009ht}, \cite{Gallego:2020vbe}.

Note that even though in Eq.~\eqref{v_total_inf} and in Eq.~\eqref{v_total_thermal}, we have written the total potential during and after inflation separately, in reality, both the  $V_{\rm inf}$ and $V_{\rm thermal}$ exist simultaneously. In fact, the former translates to the thermal bath due to the process of reheating and it causes the temperature-dependent contributions to the potential. In our work, we will consider this conversion of energy consistently and study how the process affects the dynamics of the moduli stabilization. We will emphasize that the condition of not having a minimum for $V_{\rm total}$ as discussed in \cite{Buchmuller:2004xr}, \cite{Buchmuller:2004tz} is not the same as the condition of destabilization of the potential.  In another way, the effective potential becoming run-away nature does not correspond to the destabilization of the modulus field. The dynamics of the field need to be accounted for \cite{Barreiro_2008}, \cite{Papineau:2009tv}. In fact, the question of destabilization is always initial field value dependent, and our analysis shows that the corrections due to radiation bath do not make things worse in any way.

For any value of temperature, there will always be an initial field range for which the field does not overshoot.  When $T \gtrsim T_{\rm crit}$, the local minimum is no longer there, and the effective potential is run-away nature. The issue is involved with the dynamics of the field as well as changing the form of the potential as time passes. The field is moving under the Hubble damping which is proportional to the energy density of the Universe that includes either the inflaton energy density or the radiation produced through reheating. The Hubble damping helps the field to settle at its minimum for some range of initial field values. If the initial temperature $T \lesssim T_{\rm crit}$, the potential minimum always exists. In this case, if the initial field value is larger than the $\sigma_{\rm max}$, the field overshoots. Again, there will be another field value smaller than the $\sigma_{\rm max}$ for which the field will overshoot due to attaining high initial kinetic energy at the steep part of the potential.

 If we consider radiation energy density as an initial condition \cite{Barreiro_2008}, its energy density is continuously falling and thus affecting the shape of the potential. Therefore, it is possible that before the field crosses the barrier height, an instantaneous minimum is created again and the field gets trapped without destabilizing the potential. Moreover, in this case, the position of the maximum and the minimum change with time. Surely, the question of overshooting depends on the initial field position of the modulus field. We will analyze how the initial allowed field range varies with the initial temperature. Contrary to the usual notion, we will find that the allowed field range is larger for temperature above $T_{\rm crit}$.

 Again, reheating is not an instantaneous phenomenon; the temperature is produced in a gradual manner. It allows the field to relax to its instantaneous minimum more easily than the case when the radiation bath is assumed as an initial condition. In this case, the relaxation of the modulus field to its minimum depends on the total decay width of the inflaton $\varphi$. Before appreciable decay happens, the modulus feels only the zero-temperature potential. Now, the smaller the decay width, it takes longer time to produce the thermal bath, thus distorting the potential. Broadly, the effects of reheating allow the field to stabilize for a larger range of initial conditions. Our analysis will focus on this issue in detail. We will also discuss how the allowed initial field range causes cosmological moduli problems by violating BBN bound on nucleosynthesis.

This paper is organized in the following manner. In the next section, we will discuss how thermal baths correct the zero-temperature potential, and without considering dynamics, we will calculate the critical temperature when the potential becomes of a runaway nature. In Sec. \ref{Dynamics}, we will analyze the dynamics of the field when a radiation bath is considered as an initial condition. Our focus would be to find out the allowed initial field range for different values of temperature ranging from below $T_{\rm crit}$ to above $T_{\rm crit}$. In Sec. \ref{Effect of Continuous reheating and Moduli Dynamics}, we will focus our attention on accounting for the radiation bath generation from the decay of the inflaton. We will find out how it further relaxes the allowed field range. In Sec. \ref{moduli_abundance}, we discuss moduli abundance constraints on initial field values from BBN, and we will contrast it with the constraints coming from overshooting. In Sec. \ref{Discussions and Conclusion}, we discuss and conclude.

\section{Thermal corrections and critical temperature}
\label{thermal_corrections}

At the end of inflation, a radiation bath is produced from the gauge and matter fields. In turn, the zero-temperature modulus potential  $V_0 (\sigma)$ receives temperature-dependent corrections. The corrections depend on the masses and couplings of the particles present in the thermal bath. The moduli particles typically do not take part in the thermal bath due to their Planck suppressed couplings\footnote{But, see also \cite{Anguelova:2009ht}.}. But, gauge and matter fields contribute to the modulus potential through loops.

In thermal field theory, the free-energy $F(g, T)$ contributes to the effective potential, where $g$ is the gauge coupling constant. At high temperature, the free energy has a perturbative expansion in $g$, and up to the leading order it is given by $F(g, T)=(a_0+a_2 g^2)T^4$ \cite{Kapusta:2006pm}. The parameters $a_0 (<0)$ and $a_{2} (>0)$ depending on the underlying gauge theory and the matter content considered, and for our consideration, we will treat $a_{0}$ and $a_{2}$ as free variables. The first term originates from the $1$-loop thermal corrections that represent the ideal gas of non-interacting particles. On the other hand, the second term corresponds to the interactions among the particles in the thermal bath and appears in the $2$-loop level.

In this case, the temperature-dependent effective potential looks like \begin{equation} \label{v_total}V_{\rm total} =  V_0 (\sigma) +  (a_0+a_2 g^2)T^4~. \end{equation}In String theory the gauge coupling constant is related to the modulus field: $g^2 = \kappa/\sigma$, where $\kappa$ is a constant of $\mathcal O$(1) \cite{Buchmuller:2004xr},\cite{Buchmuller:2004tz}. In the end, the temperature-dependent part of the potential becomes moduli-dependent with its runaway nature, and its strength is governed by the temperature.  The zero-temperature potential typically has a local minimum separated from its minimum at infinite vev by a finite barrier height  \cite{Kachru:2003aw},\cite{Kallosh:2004yh}. At sufficiently high temperature, the potential completely becomes of the run-away nature and the field asymptotically goes to large vev with gauge coupling becoming small. Therefore, in analyzing the dynamics of a modulus field at the end of inflation, in addition to the $V_0 (\sigma)$, the temperature-dependent part of the potential is crucial, more specifically, how the temperature is produced.

The zero-temperature potential $V_0(\sigma)$ originates from some moduli stabilization mechanism in String Theory. The potential at its minimum is positive only after the addition of an uplifting term to an otherwise supersymmetric anti-De Sitter minimum. Due to the uplifting, the SUSY is broken and a finite barrier is created whose height is almost equal to the depth of the anti-De Sitter minimum. Therefore, in this kind of set-up $V_{\rm max} \sim m_{3/2}^2 M_{\rm pl}^2$, where $m_{3/2}$ is the gravitino mass parameterizing the supersymmetry breaking scale. The critical temperature is the temperature at which the finite minimum of the moduli potential disappears. Mathematically, the critical temperature $T_{\rm crit}$ is defined by the appearance of a saddle point at some field value $\sigma_{\rm crit}$:

\begin{equation}
V_{\rm total}'(\sigma_{\rm crit},T_{\rm crit}) = 0 \label{1deri}, ~~ V_{\rm total}''(\sigma_{\rm crit},T_{\rm crit}) = 0~.
\end{equation}

Now, empirically the above conditions for the disappearance of the potential minimum translate to the condition of $V_{\rm total} \gtrsim \mathcal{O}(1) V_{\rm max}$, and in terms of the temperature, it translates to $T_{\rm crit}\sim V_{\rm max}^{1/4}$. Using the relation between the $V_{\rm max}$ and the gravitino mass, the approximate expression for the critical temperature is \cite{Buchmuller:2004tz}\begin{equation}\label{crit temp} T_{\rm crit} \sim c \sqrt{m_{3/2}M_p}~,\end{equation}where $c \sim \mathcal{O}(1)$, and it depends on the explicit model parameters \cite{Buchmuller:2004xr},\cite{Buchmuller:2004tz}.

\begin{figure}[h!]
\centering \includegraphics[scale=0.4]{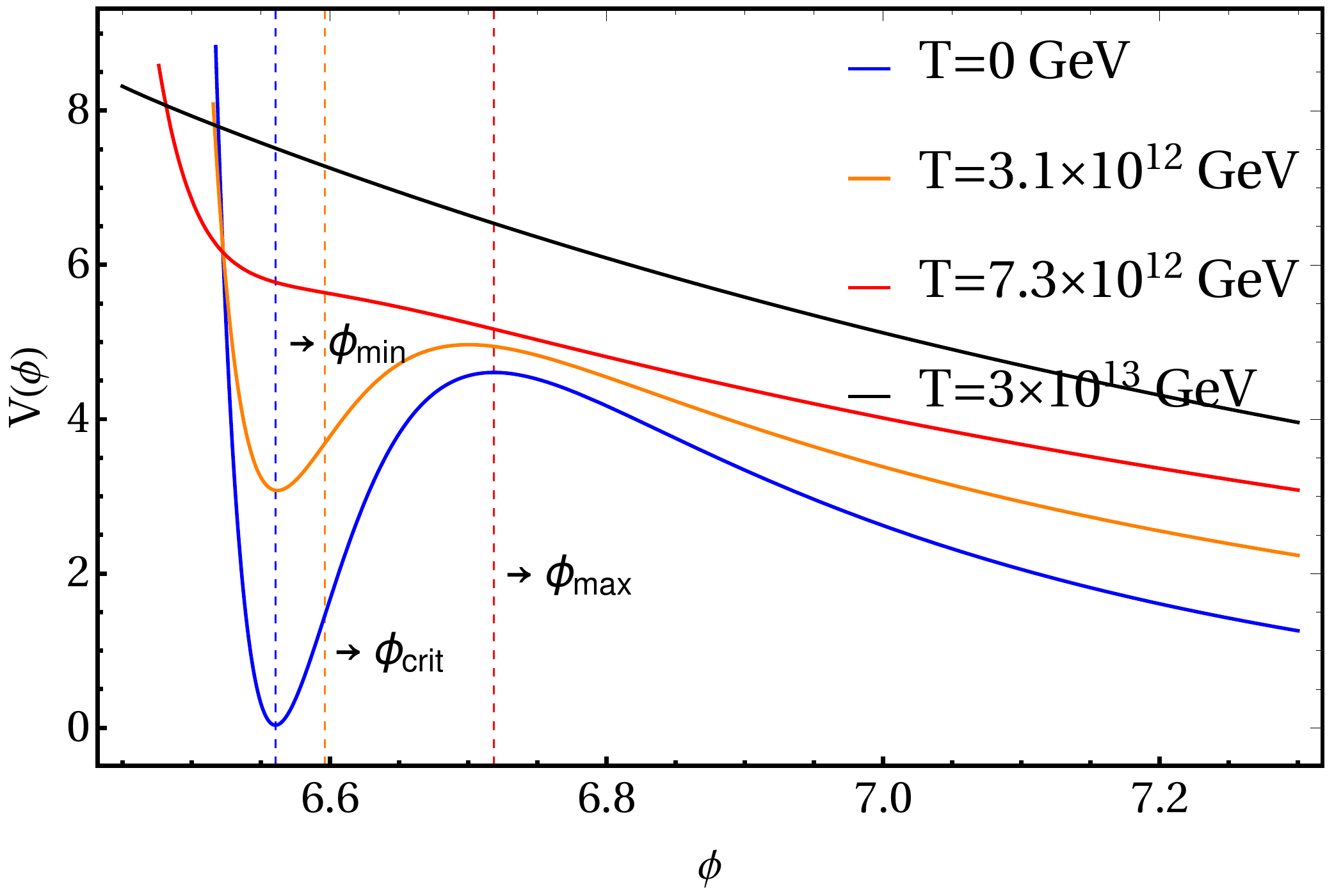} \caption{The potential of Eq.~\eqref{v_total} with $V_0(\sigma)$ given by the KKLT potential of Eq.~\eqref{pot} for $A=1, a=0.1, W_0 = -10^{-8}$ and $D= 3.3\times10^{-17}$. The vertical axis is in arbitrary scale, and the $x$-axis is the canonically normalized modulus field in Planck units. The dotted vertical lines show the position of the minimum ($\phi_{\rm min}$) and the maximum ($\phi_{\rm max}$) of the zero-temperature potential, and the critical value ($\phi_{\rm crit}$). Here  $\phi=\sqrt{3/2}\ln(\sigma)$.} \label{fig:KKLT_potential}
\end{figure}

For example, here we quickly review the KKLT moduli stabilization potential and the effects of thermal corrections on the potential. We will need the form of the potential in our future analysis. In the KKLT set-up, the dilaton and the complex structure modulus are stabilized at a higher energy scale by suitable choices of fluxes. The low-energy effective potential is governed by one K\"ahler modulus that corresponds to the overall volume of the internal space. The superpotential and the K\"ahler potential for the complex volume modulus $\rho=\sigma+i\alpha$ are given respectively by \cite{Kachru:2003aw}\begin{equation}	W=W_{0}+Ae^{-a\rho}, ~~~~	K=-3\ln(\rho+\overline{\rho})~.\end{equation}By appropriately choosing $W_0$ and $A$, we set $\alpha = 0$ (stabilised) for our analysis. The $F$-term scalar potential in $\mathcal{N}=1$ supergravity looks like \begin{equation}\label{pot}	V_0^{\rm KKLT}(\sigma) = \frac{aAe^{-a\sigma}}{2\sigma^2}\left(\frac{1}{3}aA\sigma e^{-a\sigma}+W_{0}+Ae^{-a\sigma}\right) + \frac{D}{\sigma^3}\end{equation}where the last term is added to make the AdS minimum to de Sitter minimum. Due to this uplifting term, the supersymmetry is broken, and the value of $D$ is chosen so that the potential at its local minimum is almost zero or close to the value of the present cosmological constant. 

For our case, we first choose the following values of the parameters $A=1, a=0.1, W_0 = -10^{-8}, D= 3.3\times10^{-17}$ and for these choices of parameters, $\sigma_{\rm min} \sim 211 $. Due to the uplifting potential, the potential has a finite barrier at $\sigma_{\rm max} \sim 241$, and the height is related to the gravitino mass $m_{3/2}$. For this case, $m_{\phi}$ is around $2.6\times10^{6}$ GeV, and following Eq.~\eqref{crit temp}, the corresponding  critical temperature is around $7.3\times10^{12}$ GeV . The temperature-dependent potential for these choices of parameters is shown in Fig.~\ref{fig:KKLT_potential}.  For several well-motivated models of the racetrack and K\"ahler stabilization, the critical temperature has been calculated, and it has been shown that the results are closely tied to the supersymmetry breaking \cite{Buchmuller:2004xr}. The effects of finite temperature have also been discussed in several String models \cite{Anguelova:2007ex},  \cite{Papineau:2008xf}. 

We will also consider parameter values that correspond to the moduli mass $m_{\phi} \sim 40$ GeV. This can be realised with the choice of $A=1, a=0.1, W_0 = -2.96\times10^{-13}, c= 3\times10^{-26}$. In this case, the maximum and the minimum of the potential is at $\sigma_{\rm min}=320, \sigma_{\rm max}=353$ respectively, and the corresponding critical temperature is around $10^{10}$ GeV. The above-mentioned choices of parameters that lead to the moduli masses of $m_{\phi} \sim 10^6$ GeV and $\sim 40$ GeV are for some specific reason. Irrespective of the moduli abundance, for the mass of $10^6$ GeV, the particle will decay well before the BBN. Therefore, in this case, there is no cosmological moduli problem as mentioned before. On the other hand, for the case of $m_\phi \sim 40$ GeV, the field will decay after BBN, and to avoid spoiling BBN predictions, its abundance must be suppressed adequately. As we will see later in Sec.~\ref{moduli_abundance}, the severity of moduli destabilization will crucially depend on the moduli masses. 
	
For all the above examples, the potential has one minimum which got uplifted to the supersymmetry breaking the Minkowski minimum. But, in certain models, the modulus can be stabilized at the supersymmetric Minkowski minimum whose barrier height is unrelated to the $m_{3/2}$. On the other hand, $T_{\rm crit}$ is related to the barrier height. In this kind of set-up, therefore, $T_{\rm crit}$ can not be related to the  $m_{3/2}$. One example of this type is the superpotential in KL model \cite{Kallosh:2004yh},\begin{equation}	W=W_{0}+Ae^{-a\rho}+Be^{-b\rho},\end{equation}and K$\ddot{a}$hler potential as like the KKLT case. If we consider only real part of the field, the effective scalar potential for KL model is,\begin{equation}\label{pot1}V_{\rm KL}(\sigma)=\frac{e^{-2(a+b)\sigma}}{6\sigma^2}\left(bBe^{a\sigma}+aAe^{b\sigma}\right)\times\left[Be^{a\sigma}(3+b\sigma)+e^{b\sigma}\left(A(3+a\sigma)+3e^{a\sigma}W_{0}\right)\right],\end{equation}For the choice of parameters $A=1, B=-1.03, a= \frac{2\pi}{100}, b=  \frac{2\pi}{99}, W_{0}=-2\times10^{-4}$ the potential has minimum at finite position with zero potential  value \cite{Kallosh:2004yh}. In this case, we do not need any supersymmetry breaking to get zero potential value at the minimum. So, $m_{3/2}$ is equal to zero, and the Eq.~\eqref{crit temp} can not be used to find out critical temperature, and it needs to be evaluated numerically. 

\begin{figure}[h!]\centering\begin{subfigure}{0.47\textwidth}\centering\includegraphics[width=\textwidth]{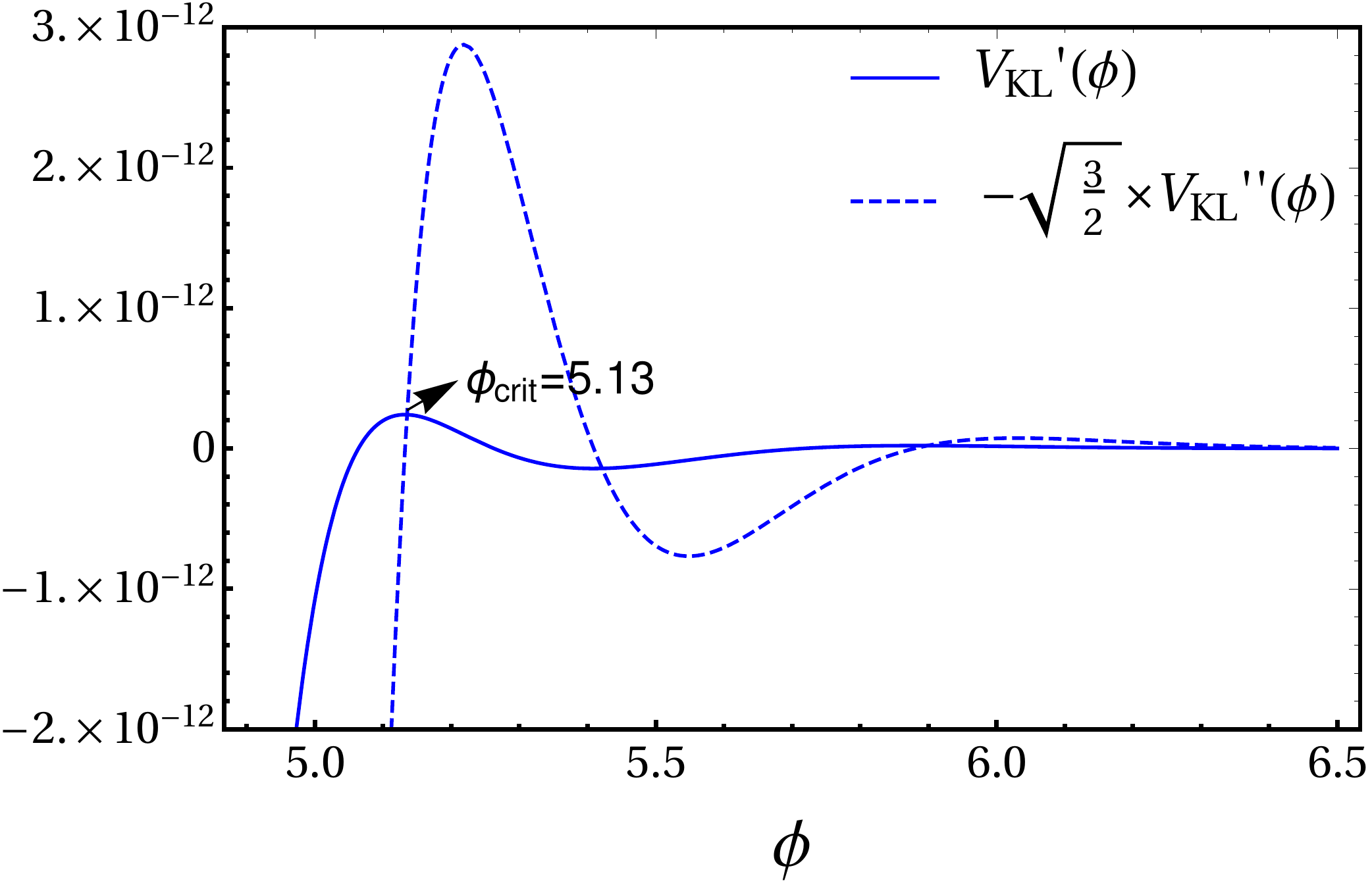}\end{subfigure}\hfill\begin{subfigure}{0.47\textwidth}\centering\includegraphics[width=\textwidth]{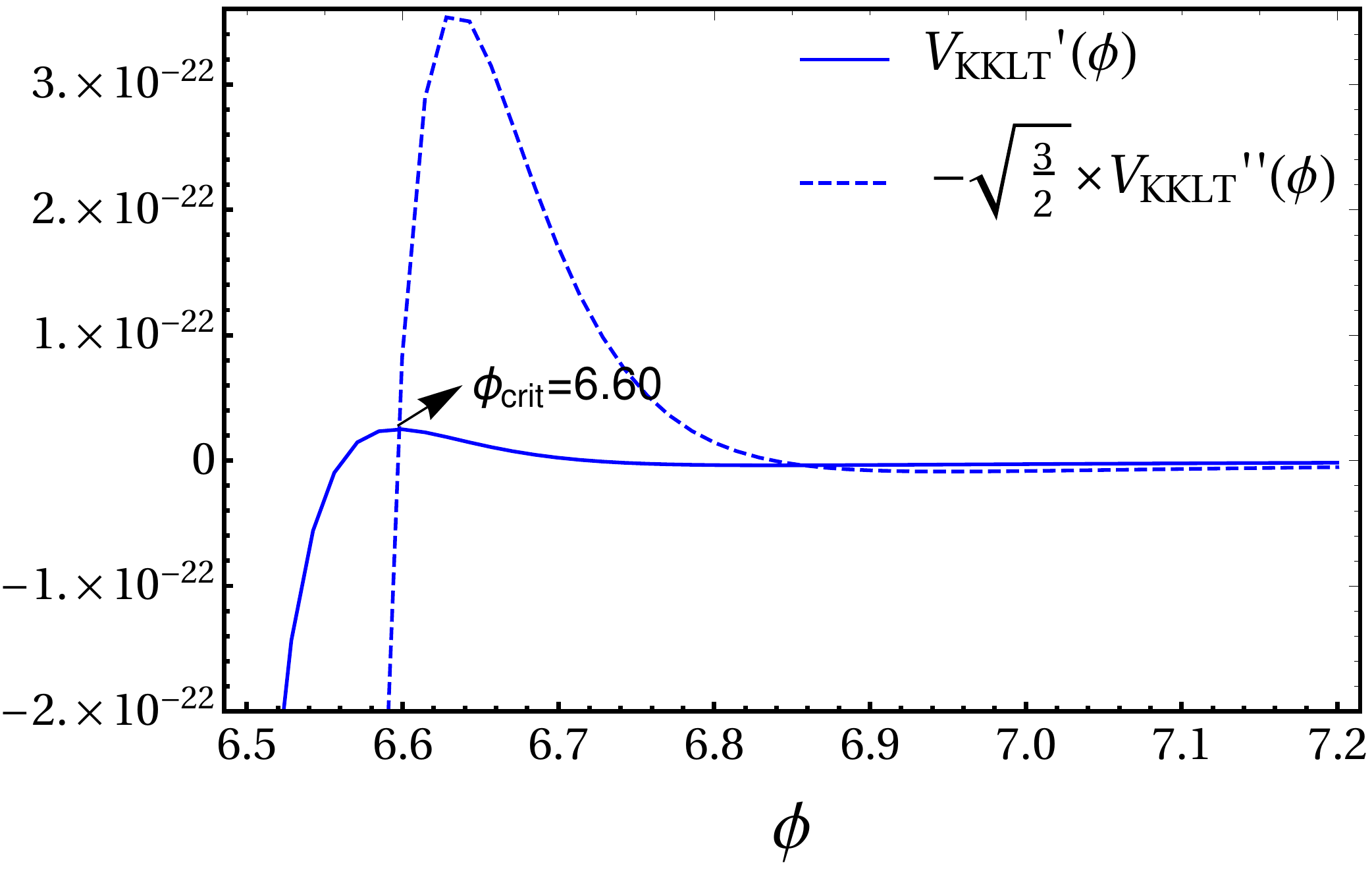}\end{subfigure}
\caption{Plots for calculating the critical field value (canonical) of moduli. The left side plot is for KL potential of Eq.~\eqref{pot1} with parameter values $A=1, B=-1.03, a=\frac{2\pi}{100}, b=\frac{2\pi}{99}, W_{0}=-2\times10^{-4}$ and the right side plot is for KKLT potential of Eq.~\eqref{pot} with parameter values  $A=1, a=0.1, W_0 = -10^{-8}, D= 3.3\times10^{-17}$.}\label{crit_field_value}\end{figure}

Now, we use the definition of the critical temperature of Eq.~$\eqref{1deri}$ for general moduli potential. Using Eq.~\eqref{1deri} for the moduli potential of Eq.~\eqref{v_total}, we obtain,\begin{equation}V'_{0}(\sigma_{\rm crit}) = \frac{a_{2}}{\sigma^2_{\rm crit}}T^4_{\rm crit}, ~~V''_{0}(\sigma_{\rm crit})=-\frac{2a_{2}}{\sigma^3_{\rm crit}}T^4_{\rm crit}\label{dri}~,\end{equation}and it leads to\begin{equation}\label{trans}\frac{V'_{0}(\sigma_{\rm crit})}{V''_{0}(\sigma_{\rm crit})}=-\frac{\sigma_{\rm crit}}{2}~.\end{equation}Here, $V_{0}(\sigma _{\rm crit})$ may be KKLT or KL potential at zero temperature. The value of $\sigma_{\rm crit}$ can be found by solving Eq.~\eqref{trans} numerically in the graphical method, and for KKLT and KL potential, we show the results in Fig.~ \ref{crit_field_value}. The $\sigma_{\rm crit}$ is the intersection of the graphs $V'_{0}(\sigma)$ and $-\left(\sigma/2\right)V''_{0}(\sigma)$.

In Fig. \ref{crit_field_value}, we have plotted the canonical modulus field $\phi=\sqrt{3/2}\ln(\sigma)$. In the left panel of Fig. \ref{crit_field_value}, we observe that there exist two points of intersections for KL potential but only one of them is valid for which $T_{\rm crit}$ is a real positive number. This happens for $\phi_{\rm crit}=5.13$ or $\sigma_{\rm crit}\simeq 65.93$, then from Eq.~\eqref{dri} we get,\begin{equation}T^{\rm KL}_{\rm crit}\simeq 5\times 10^{15}{\rm GeV},\end{equation}where $a_2$ is considered equal to 1. In the right panel of Fig.~\ref{crit_field_value} for KKLT potential, we observe that there exists only one intersection point  with parameter values $A=1, a=0.1, W_{0}=-10^{-8}, D=3.3\times 10^{-17}$ and the value of $\phi_{\rm crit}=6.60$ or $\sigma_{\rm crit}=218.96$, then from Eq.~\eqref{dri} we get,\begin{equation}T^{\rm KKLT}_{\rm crit}\simeq 2\times 10^{13}~ {\text GeV}~,\end{equation}which is almost equal to the critical temperature calculated using the Eq.~\eqref{crit temp}.

The above analysis of critical temperature has the following drawbacks. Firstly, the analysis is only related to finding the inflection point of the effective temperature-dependent potential. Reaching the associated temperature does not necessarily mean that the moduli are destabilized. As mentioned earlier, the moduli will be destabilized only when the field crosses the top of the barrier height, and this is a dynamical question that necessarily depends on the initial field values. The dynamical analysis involves Hubble damping which includes any background energy density present in the system. Moreover, due to the non-canonical nature of the kinetic term, analysis of the effective potential is not sufficient.

Secondly, in any realistic set-up, the radiation bath needs to be produced at the end of inflation. In the standard cold inflation scenario, the Universe is solely dominated by the inflaton energy density at the end of inflation. Now, the production of temperature is a continuous process with its associated time scale. It is crucial to understand how the field evolves while the reheating temperature of the thermal bath is produced. The most well-understood source of the thermal bath is due to the decay of inflaton via the process of reheating. Therefore, it is crucial to analyze the dynamics of the field in the expanding background where the production of thermal bath is also taken care of. In the next sections, we will discuss these effects, and see how it affects the range of initial conditions for successful moduli stabilization. 

\section{Dynamics with initial radiation bath}
\label{Dynamics}
In this section, we will discuss the dynamics of a modulus field in the presence of a thermal bath. For now, we will not worry about how this thermal bath is created, and therefore the radiation energy density will be considered as an initial condition  \cite{Barreiro_2008}. To simplify our discussions, we will consider the dynamics of the real part ($\sigma$) of the complex field $\rho=\sigma+i\alpha$. A more general two-field analysis with two initial temperatures can be found in  \cite{Barreiro_2008}. In contrast to the previous analysis, we will study how the allowed field space is changed when the initial temperature is varied in a suitable range.

The dynamics of the modulus field, in this case, is governed by the following equations,

\begin{align}
&\ddot{\sigma}+3H\dot{\sigma}-\frac{\dot{\sigma}^2}{\sigma}+\frac{2\sigma^2}{3}V'_{\rm total}(\sigma,\rho_{r})=0 \label{dyn1}\,, \\
&\dot{\rho_{r}}+4H\rho_{r}=0 \label{radiation_1}\,,\\
&3M^2_{p}H^2=\frac{3}{4}\left(\frac{\dot{\sigma}}{\sigma}\right)^2+V_{0}(\sigma)+\rho_{r} \label{energy1}~.
\end{align}
The Eq.~\eqref{dyn1} is the modulus field evolution where the third and the fourth terms also get contributions from the non-canonical K\"ahler potential. Here $\rho_{r}$ is the background radiation energy density that evolves with time in the expanding spacetime via Eq.~\eqref{radiation_1}, and the last Eq.~\eqref{energy1} is the Friedmann equation. We can obtain the pressure and the energy density of the thermal fluid as $P_{r}=F(g,T)$ and $\rho_{r}=-P_{r}+T\frac{dP_{r}}{dT}$, and hence,

\begin{equation}
\rho_{r}=-3a_{0}(1+r g^2)T^4 , \label{energy density}
\end{equation}
where $r=a_2/a_0$, $g$ is the gauge coupling constant, and $a_2$ and $a_0$ depend on the micro-physics of the thermal bath.  The Eq.~\eqref{energy density} tells us the relation between the temperature and the radiation energy density and we use either of these quantities interchangeably.   So, the temperature corrected potential of Eq.~\eqref{v_total} in terms of $\rho_{r}$  can be rewritten as,

\begin{equation}\label{eff_pot}
V_{\rm total}=V_{0}(\sigma)-\frac{1}{3}\frac{r\rho_{r}g^2}{1+r g^2} +a_{0}T^{4}~.
\end{equation}
For our consideration $V_0 (\sigma)$ would be the KKLT potential for two specific choices of parameter sets given in Sec.~\ref{thermal_corrections}, and $g^2 = \kappa/\sigma$ with $\kappa$ being $4\pi$.

 It has been noted earlier that when the initial temperature $T_{\rm init}$ related to the radiation bath is larger than the critical temperature $T_{\rm crit}$, the effective potential is without a minimum. But, as time progresses the effects of the temperature-dependent part of the potential decrease, and the local minimum and the maximum (barrier height) start to appear, eventually going to the form of the zero-temperature potential. In this situation when $T_{\rm init} > T_{\rm crit}$, the field does not necessarily overshoot, and it becomes dependent on the initial field values. If the field starts to move from the far left, the field gains enough kinetic energy to cross the barrier even if the minimum is created before the overshooting time. We denote this value from the left side by $\phi^{\rm L}_{\rm init}$ for which the field just overshoots.

	 Similarly, there will be another field value larger than $\phi^{\rm L}_{\rm init}$ for which field again overshoots and we denote that by $\phi^{\rm R}_{\rm init}$. We will see soon that the instantaneous position of the maximum (minimum) of the potential moves toward larger (smaller) values as the temperature decreases. Therefore, the $\phi^{\rm R}_{\rm init}$ will be always smaller than the field value at which zero temperature potential has the maximum. When the initial temperature is smaller than the critical temperature, the minimum of the potential exists from the very beginning. In summary, for field values between $\phi^{\rm L}_{\rm init}$ and $\phi^{\rm R}_{\rm init}$, the modulus field does not overshoot, and we can have consistent cosmology as long as the modulus field satisfies other cosmological bounds related to its abundance.

To solve the dynamics of the field, we will always take the initial field velocity to be zero. For some reasons if the initial field velocity is large, allowed field space for not overshooting the barrier will shrink. We will take $r = -1.3 $ for our initial analysis, and at the end, we will discuss the effects of varying $r$. 
		\begin{figure}[h]
		\centering
		\hspace{-0.9cm} 
		\includegraphics[width=.5\textwidth]{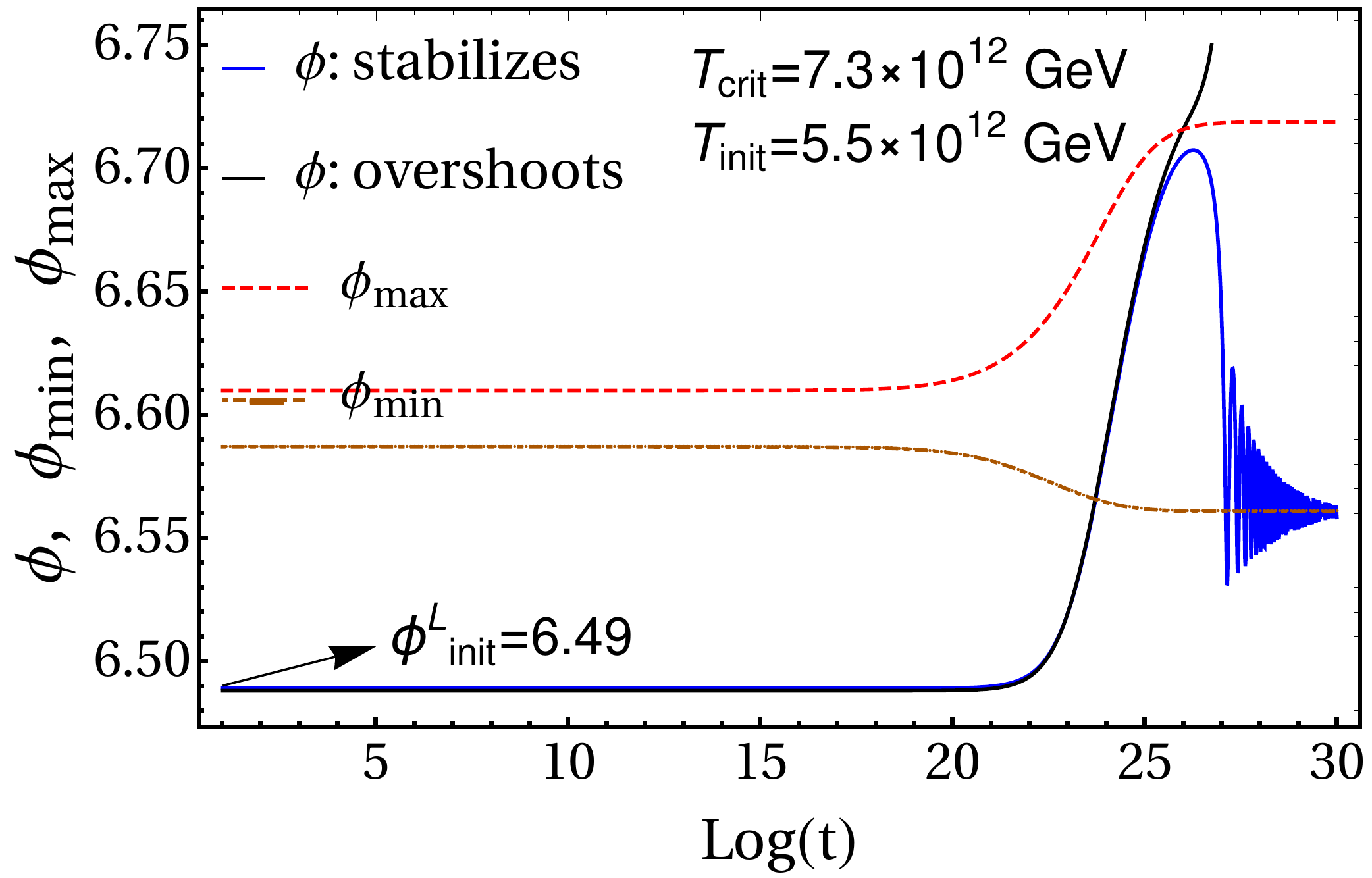}
		\hspace{.2cm}
		\includegraphics[width=0.5\textwidth]{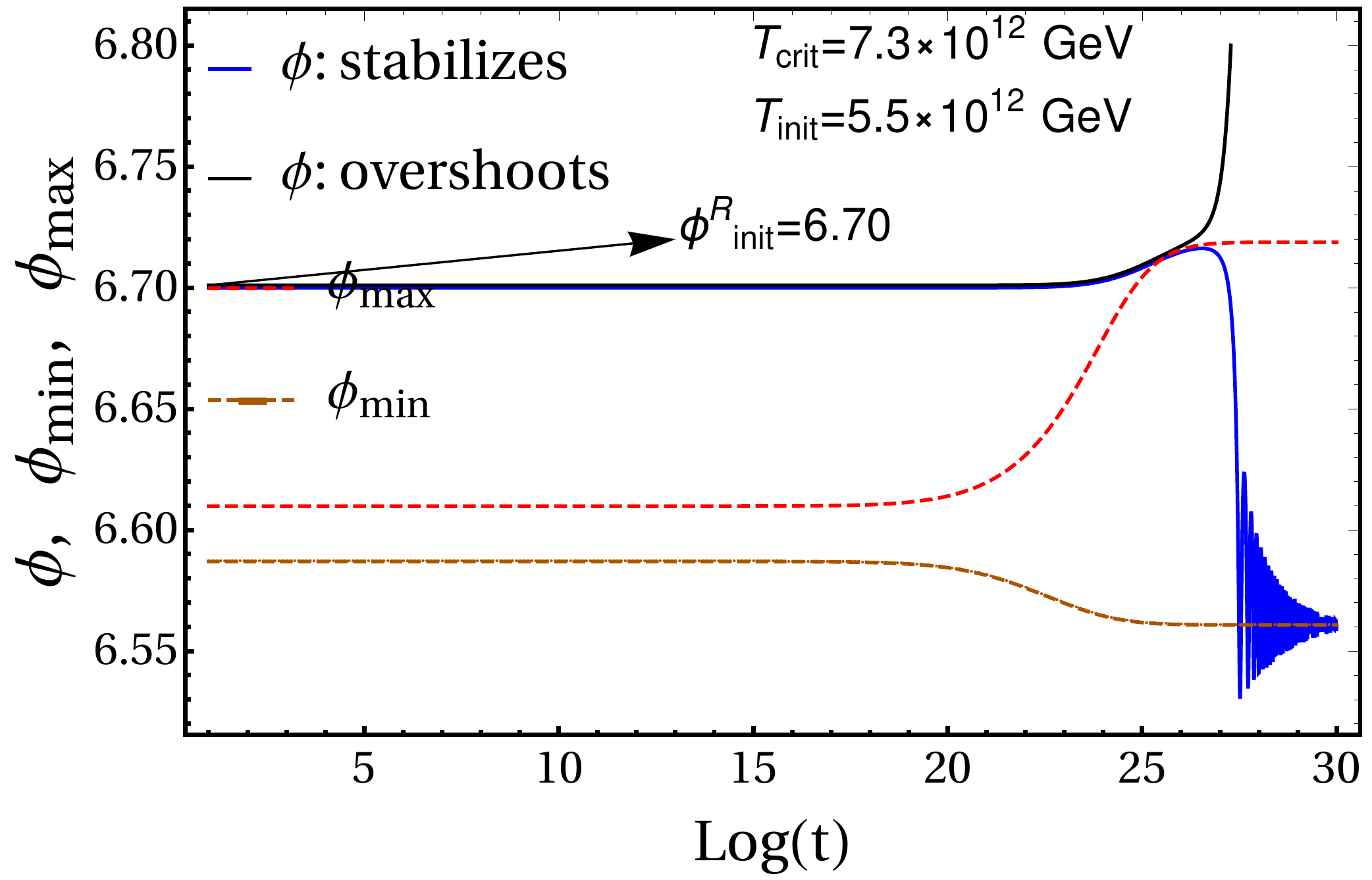}
		\caption{Time evolution of the canonical modulus field is shown where the blue line corresponds to stabilization, and the black one for overshooting. The dashed red and dot-dashed brown lines show the instantaneous positions of the maximum and minimum of the potential, respectively. The plots are for initial temperature smaller than the critical temperature. The left panel is for $\phi_{\rm init} < \phi_{\rm min}$ and the right panel is for $\phi_{\rm init} > \phi_{\rm min}$. The initial field values $\phi^{\rm L}_{\rm init}$ and $\phi^{\rm R}_{\rm init}$, for which overshooting just does not happen, are also marked in the plot.}
		\label{dynamics1}
	\end{figure}
	
We first discuss the dynamics of the modulus field for the case of critical temperature around $7.3\times10^{12}$ GeV; see Fig.~\ref{fig:KKLT_potential} for the form of the effective potential.  When the initial temperature is below, but close to the critical temperature, the evolution of the field is shown in the panels of Fig. \ref{dynamics1}. 
In these plots, the dashed red line corresponds to the instantaneous position of the maximum, and the dot-dashed brown line represents the instantaneous position of the minimum. Note that the positions of the minimum and the maximum are closer while the temperature is large, and as the temperature drops down, the maximum moves to higher field values, and the minimum moves to lower field values with asymptotic values being for the case of zero temperature potential. The blue line corresponds to the case for which the initial field value is such that the field does not overshoot, and in the end, it oscillates around the minimum. But, in the case of the black line the field just overshoots and eventually reaches the infinite vev representing destabilization of the field. 

The left panel of  Fig. \ref{dynamics1} shows the dynamics when the field starts to move with $\phi_{\rm init} < \phi_{\rm min}$, and in this case, as soon as the field crosses the barrier with finite velocity, the field overshoots. The right panel shows the case $\phi_{\rm init} > \phi_{\rm min}$, and in this case, the initial field value is even greater than the instantaneous maximum of the potential, i.e the field is at the right side of the barrier. The Hubble damping due to high initial temperature with initial zero velocity holds the field at its place. With time, the instantaneous maximum moves towards larger field values, and eventually the field becomes on the left side of the barrier. It means that if the field crosses the instantaneous barrier, that does not necessarily mean destabilization. In summary, we see that the field does not overshoot the barrier height if the initial field value is within the range $6.5 < \phi_{\rm init}< 6.7$, and obviously this range varies depending on the initial temperature. At the end of this section, we will show the variations of the allowed range with the initial temperature.  If the initial temperature is much lower than the critical temperature, the changes in the instantaneous minimum and the maximum are not appreciable. In this case, the value of
$\phi^{\rm R}_{\rm init}$ is governed by the position of the barrier i.e if the initial field value is greater than the $\phi_{\rm max}$, the field immediately overshoots. On the other hand, the value of $\phi^{\rm L}_{\rm init}$ is fully determined by the slope of the effective potential.

\begin{figure}[h]

	\centering



	\hspace{-.9cm} \includegraphics[width=.5\textwidth]{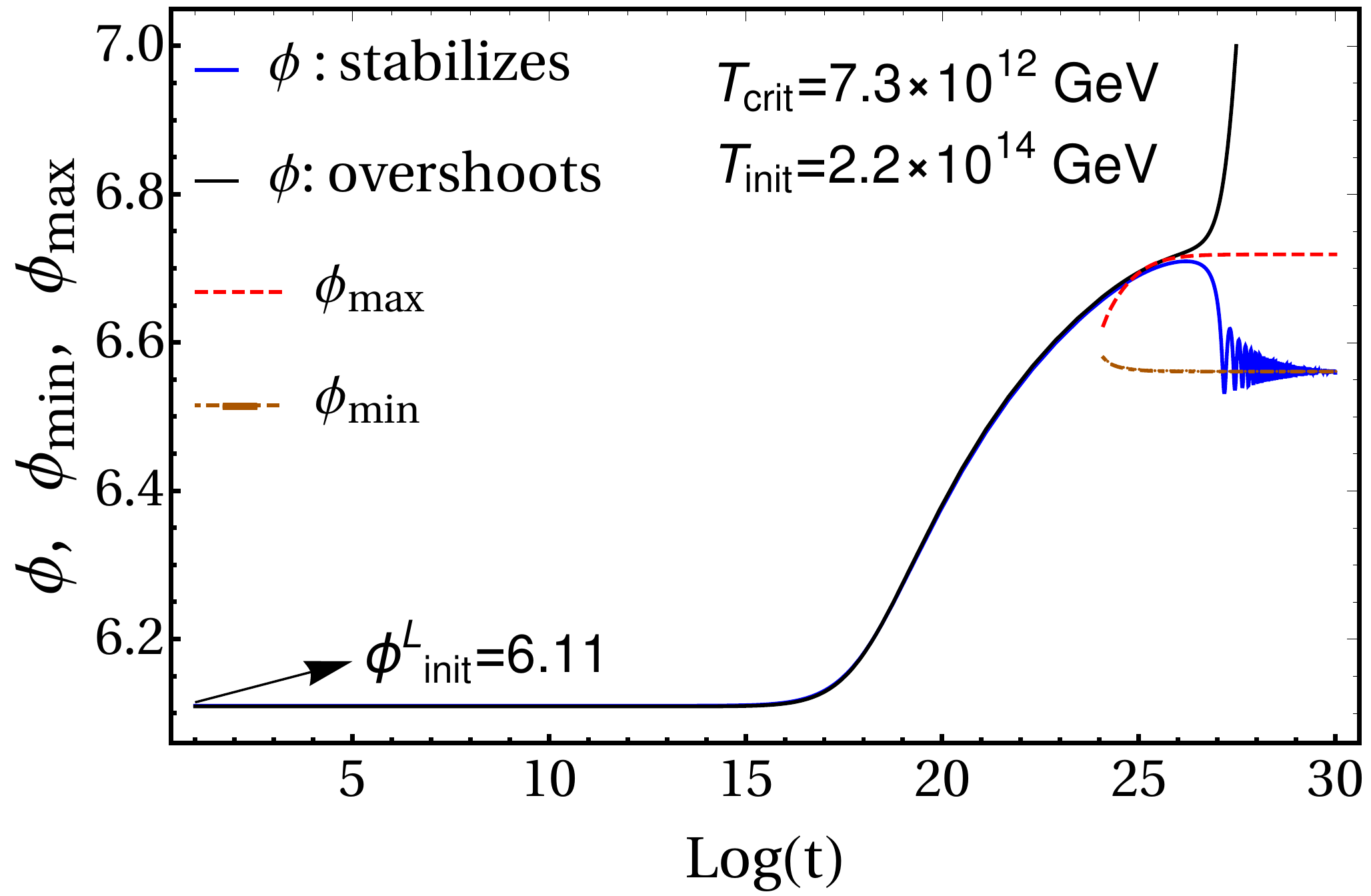}
	\hspace{.2cm}
	\includegraphics[width=0.5\textwidth]{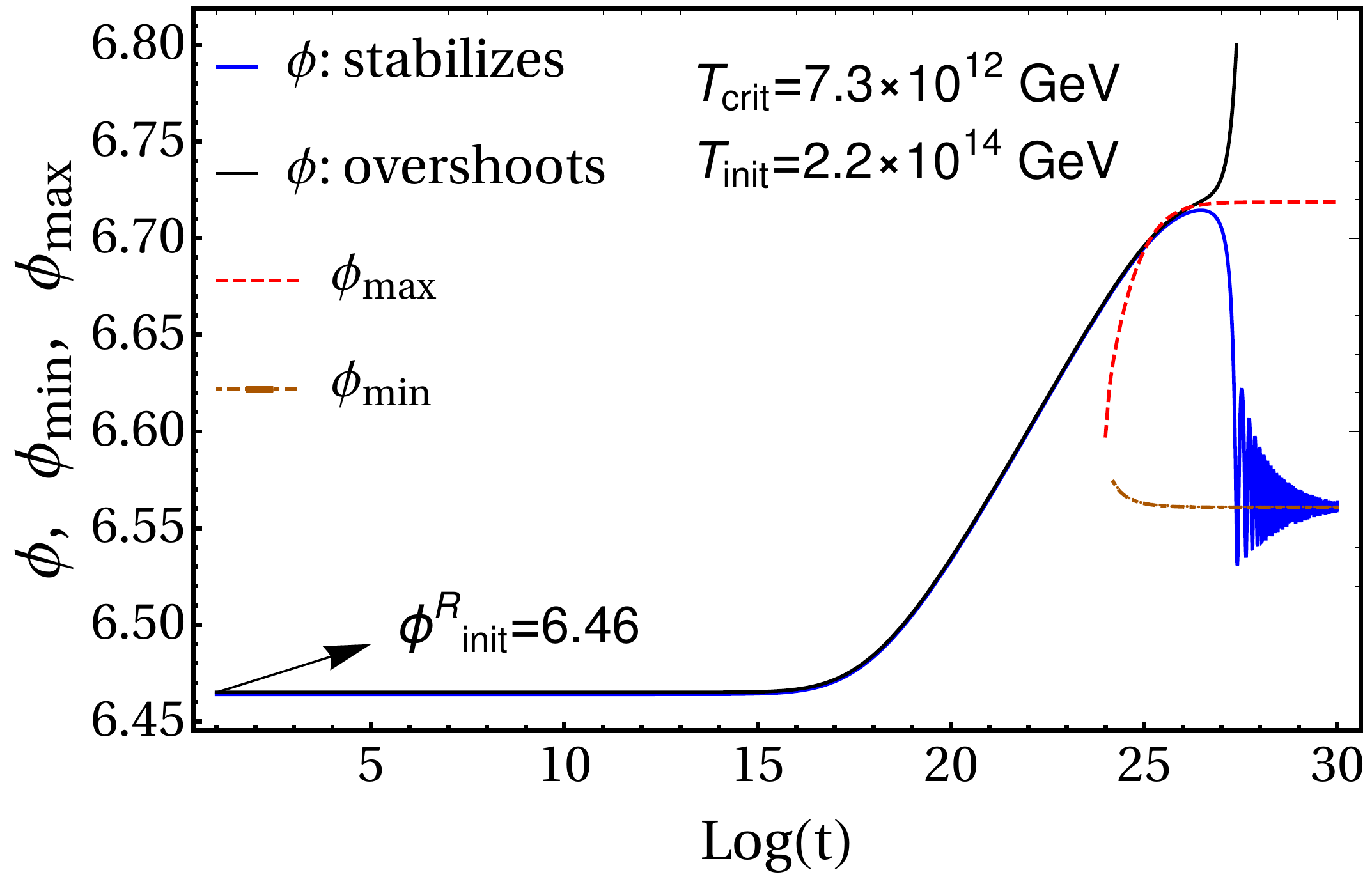}
	\caption{ As like Fig.~\ref{dynamics1}, but with an initial temperature being greater than the critical temperature. The dotted lines appear when the temperature becomes low enough such that the effective potential shows a minimum and a barrier.}
	\label{dynamics3}
\end{figure}

In Fig.~\ref{dynamics3}, the dynamics of the field are plotted when the initial temperature is much above the $T_{\rm crit}$. In this case, the effective potential is run-away nature to start with, and the field does not necessarily overshoot, and it depends on the initial conditions of the field. In fact, as is seen from the plot, the dashed red line  (instantaneous maximum) line and the dot-dashed brown line (instantaneous minimum) exist only after a certain time. It is evident that as soon as the field crosses the instantaneous maximum, the field overshoots to larger values. For the particular choice of initial temperature, both  $\phi^{\rm L}_{\rm init}$  and  $\phi^{\rm R}_{\rm init}$ are on the two sides of the instantaneous minimum vev. But, for larger initial temperatures, it turns out that both the initial field values are smaller than the $\phi_{\rm min}$.

\begin{figure}[h]
		\centering
		\hspace{-0.8cm} \includegraphics[width=.49\textwidth]{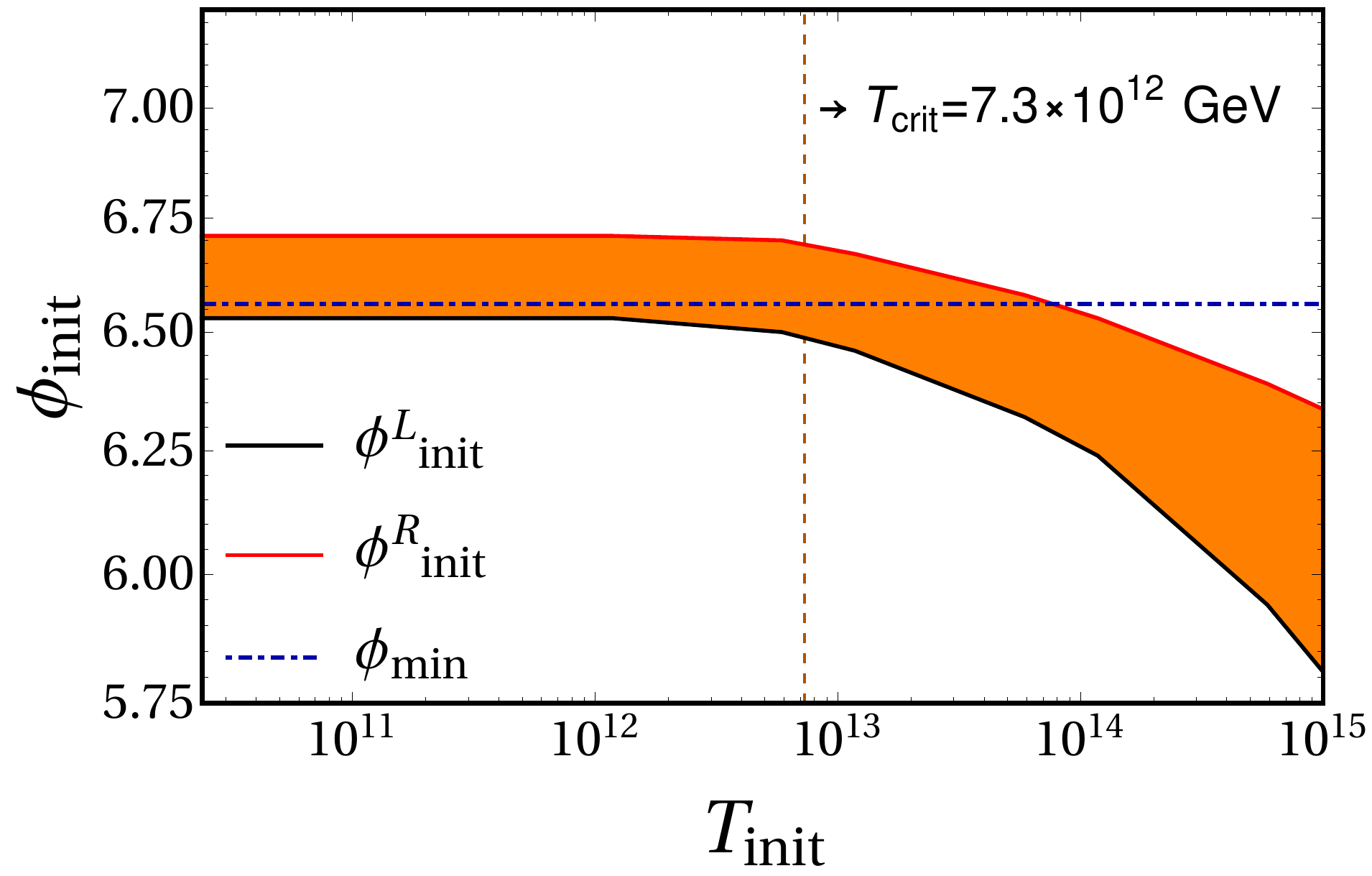}
		\hspace{.5cm}
		\includegraphics[width=0.49\textwidth]{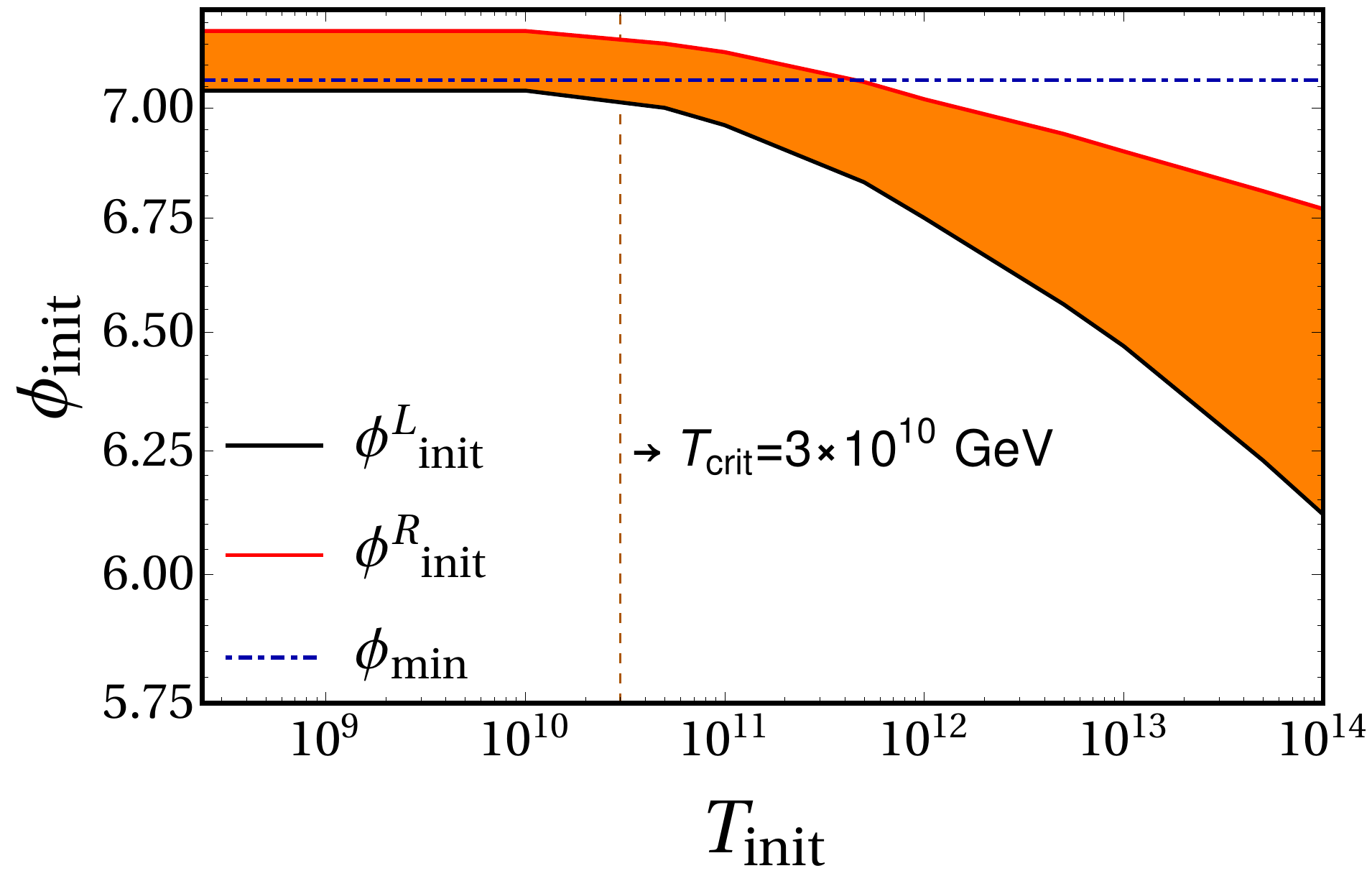}
		\caption{Plots of the initial field range vs initial temperature of the radiation bath. The left figure for $T_{\rm crit} \sim 10^{12}$ GeV. The right figure is for for $T_{\rm crit} \sim 10^{10}$ GeV. The orange dashed line refers to the critical temperature for both plots. The red solid line shows $\phi^{\rm R}_{\rm init}$, and the black solid line corresponds to $\phi^{\rm L}_{\rm init}$. The dot-dashed blue line shows the position of the zero-temperature minimum.}
		\label{dynamics4}
	\end{figure}

Finally, in Fig.~\ref{dynamics4} (left panel), we show how the allowed initial field range changes with the temperature of the radiation bath. When the temperature is much smaller than the $T_{\rm crit}$ (vertical orange dotted line), the instantaneous minimum and the maximum nearly overlap with their zero temperature values respectively. In this case,  $\phi^{\rm R}_{\rm init}$  is determined by the position of the maximum that does not change appreciably with temperature as long as $T_{\rm init} \ll T_{\rm crit}$. It makes  $\phi^{\rm R}_{\rm init}$ nearly independent of temperature below $T_{\rm crit}$. On the other hand, for $T_{\rm init} \ll T_{\rm crit}$ the potential becomes very steep on the left side of the minimum, and the initial condition $\phi^{\rm L}_{\rm init}$ is fully determined by whether the gained kinetic energy is enough to overshoot the barrier. Again, this becomes effectively independent of temperature.  This explains why the field range is insensitive to initial temperature as long as it is sufficiently smaller that $T_{\rm crit}$.

On the other hand, when $T_{\rm init}$ becomes comparable or larger than the $T_{\rm crit}$, the dynamics of the field are fully governed by the temperature-dependent part of the potential. In this case, both the values of  $\phi^{\rm L}_{\rm init}$ and  $\phi^{\rm R}_{\rm init}$ become smaller as $T_{\rm init}$ becomes larger. But, the allowed field range $\Delta \phi = \phi^{\rm R}_{\rm init} - \phi^{\rm L}_{\rm init}$ becomes larger compared to the values for $T_{\rm init} \ll T_{\rm crit}$. In fact, in this case, both $\phi^{\rm L}_{\rm init}$ and  $\phi^{\rm R}_{\rm init}$ become eventually smaller than the value of the minimum for the zero-temperature potential shown by a horizontal dot-dashed blue line.   In summary, even though for $T_{\rm init} \gg T_{\rm crit}$, the potential becomes run-away nature, the allowed field space for which no overshooting happens is large. We would like to emphasize that the issue of overshooting even exists for the zero-temperature potential. When we consider the effects of radiation bath, the issue does not become worse. In fact, the allowed field range becomes larger. In this sense, the effects of the temperature do not make the situation worse in any sense. This is one of the important conclusions we make.

 In the right panel of Fig.~\ref{dynamics4}, we again show the allowed field range when the critical temperature is smaller than the previous case, for example $T_{\rm crit} = 3\times 10^{10} \text{GeV}$. We broadly conclude that the allowed field range remains roughly the same. Obviously, for this case, the allowed range is around the zero-temperature minimum which is at a higher field value. Also for $T_{\rm init} \ll T_{\rm crit}$, the allowed range is slightly smaller due to the shorter barrier height.

\section{Reheating and moduli dynamics}
\label{Effect of Continuous reheating and Moduli Dynamics}

At the end of inflation, the radiation bath is created from the decay of the inflaton, and the process might be complicated, as well as, it will depend on the details of the model. For our analysis, we consider that the inflaton decays via perturbative processes with total decay width $\Gamma_{\varphi}$. The details of the decay process or decay products do not affect our discussions. The crucial point is that at the end of inflation, the moduli potential is still dictated by the zero-temperature potential, and as the energy density of the thermal bath starts to grow, the field starts to feel the temperature-corrected potential. The process of creating the thermal bath has its associated time scale of $\Gamma^{-1}_{\varphi}$, and that allows the modulus field time to settle around its minimum. This effect relaxes the overshooting problem. In this section, we will analyze the process in detail, and compare it to the case when the radiation bath is assumed to exist from the beginning of modulus evolution. 

At the bottom of the inflation potential where the decay process is happening during the oscillations of the field, the potential can be approximated as
\begin{equation}
V(\varphi)=\lambda\frac{\varphi^k}{M_{p}^{k-4}},
\label{inf_pot}
\end{equation}
where $\varphi$ is the inflaton field, and $\lambda$ is dimensionless coupling constant. We will consider cases of $k = 2$, or $4$. The equation of motion of the inflaton field when we include the effects of the inflaton decay can be written as,
\begin{equation}
\ddot{\varphi}+(3H+\Gamma_\varphi)\dot{\varphi}+V'(\varphi)=0,
\end{equation}
where $\Gamma_\varphi$ is the total inflaton decay width. If we assume that the decay of the inflaton is relatively slow, i.e. the oscillation time-scale is much shorter than $\Gamma_\varphi^{-1}$ and $H^{-1}$, then the governing equation for the energy density of the inflaton can be written as \cite{mukhanov:2005sc}
\begin{equation}
\dot{\rho_{\varphi}}+3H(1+\omega_{\varphi})\rho_{\phi} \simeq -\Gamma_{\varphi}(1+\omega_{\varphi})\rho_{\varphi}, \label{inflaton}
\end{equation}
where the equation of state parameter $\omega_{\varphi} = (k-2)/(k+2)$. 
The evolution of the radiation energy density produced by inflaton decay is governed by
\begin{equation}
\dot{\rho_{r}}+4H\rho_{r}\simeq (1+\omega_{\varphi})\Gamma_{\varphi}\rho_{\varphi}~. \label{radiation1}
\end{equation}
We have assumed that the system thermalises instantaneously. If the modulus field is present within this thermal bath, the dynamics of the field is dictated by 
\begin{align}
&\ddot{\sigma}+3H\dot{\sigma}-\frac{\dot{\sigma}^2}{\sigma}+\frac{2\sigma^2}{3}V'_{\rm total}(\sigma,\rho_{r})=0 \label{dyn2}\,, \\
&3M^2_{p}H^2=\frac{3}{4}\left(\frac{\dot{\sigma}}{\sigma}\right)^2+V_{0}(\sigma)+\rho_{r}+\rho_{\varphi} \label{energy2}~.
\end{align}
We will solve these Eqs. \eqref{inflaton}, \eqref{radiation1}, \eqref{dyn2} and \eqref{energy2} simultaneously and will find the initial field range for which the modulus field does not overshoot the barrier height.

 The decay process is nearly complete by the time $\Gamma^{-1}_{\varphi}$ and the decay products are thermalized with a temperature. We call that the reheating temperature $T_{\rm R}$. But, during the process of decay, the maximal temperature of the decay products is $T_{\max} \simeq \left (\Gamma_{\varphi} H_{\rm inf} M_{p}^2  \right)^{1/4}$, and it is larger than the final thermalised temperature $T_{\rm R}$ \cite{Kolb:1990vq}. Here, $H_{\rm inf}$ is the scale of inflation. Without dynamical analysis, the potential should destabilise as soon as $T_{\rm max} > T_{\rm crit}$  \cite{Buchmuller:2004xr},\cite{Buchmuller:2004tz}. Obviously, that is not the case as the field at the end of inflation feels only the zero-temperature potential. Once the decay starts to happen, the temperature bath is created with its associated distortion of the potential due to its temperature-dependent corrections. The correction is maximum at the temperature $T_{\rm max}$. In contrast to the analysis in the previous section where radiation bath is considered as an initial condition \cite{Barreiro_2008}, in this case, the radiation bath is created with the associated time-scale. In this case, the field experiences temperature-dependent corrections that are initially zero, and then gets the maximum effects at $T_{\rm max}$, and eventually again without any effect. Moreover, the produced temperature is dependent on the scale of inflation $H_{\inf}$ and the decay width of the inflaton $\Gamma_\varphi$. In addition to that, the effects depend on the parameters $k$ of Eq.~\eqref{inf_pot}, and $r$, parameterizing the effects of the gauge coupling constants, see Eq.~\eqref{energy density}. In the following discussions, we will explore dependencies on all these parameters.
\begin{figure}[h]
\centering
\hspace{-0.8cm} \includegraphics[width=.49\textwidth]{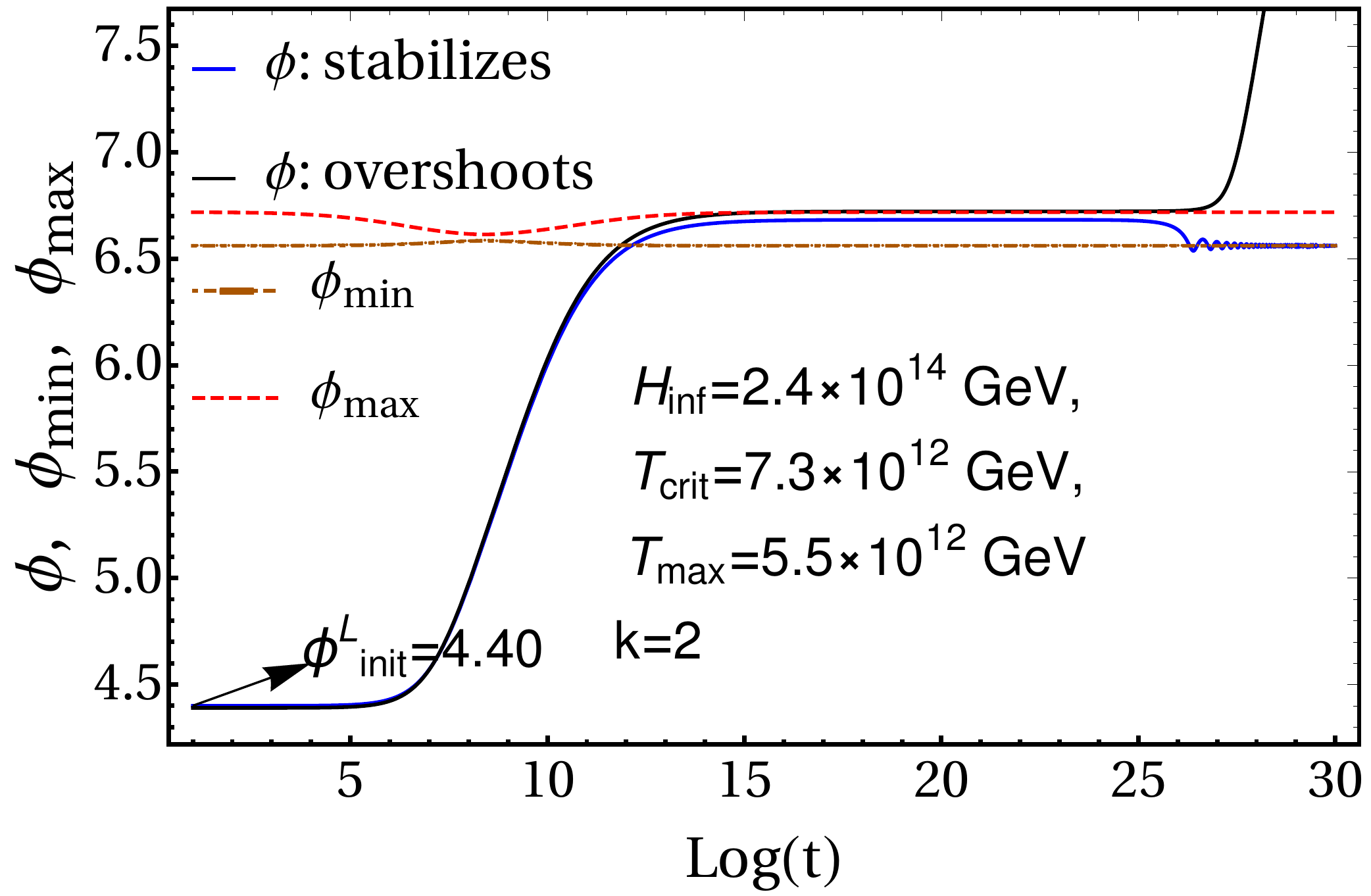}
\hspace{.5cm}
\includegraphics[width=0.49\textwidth]{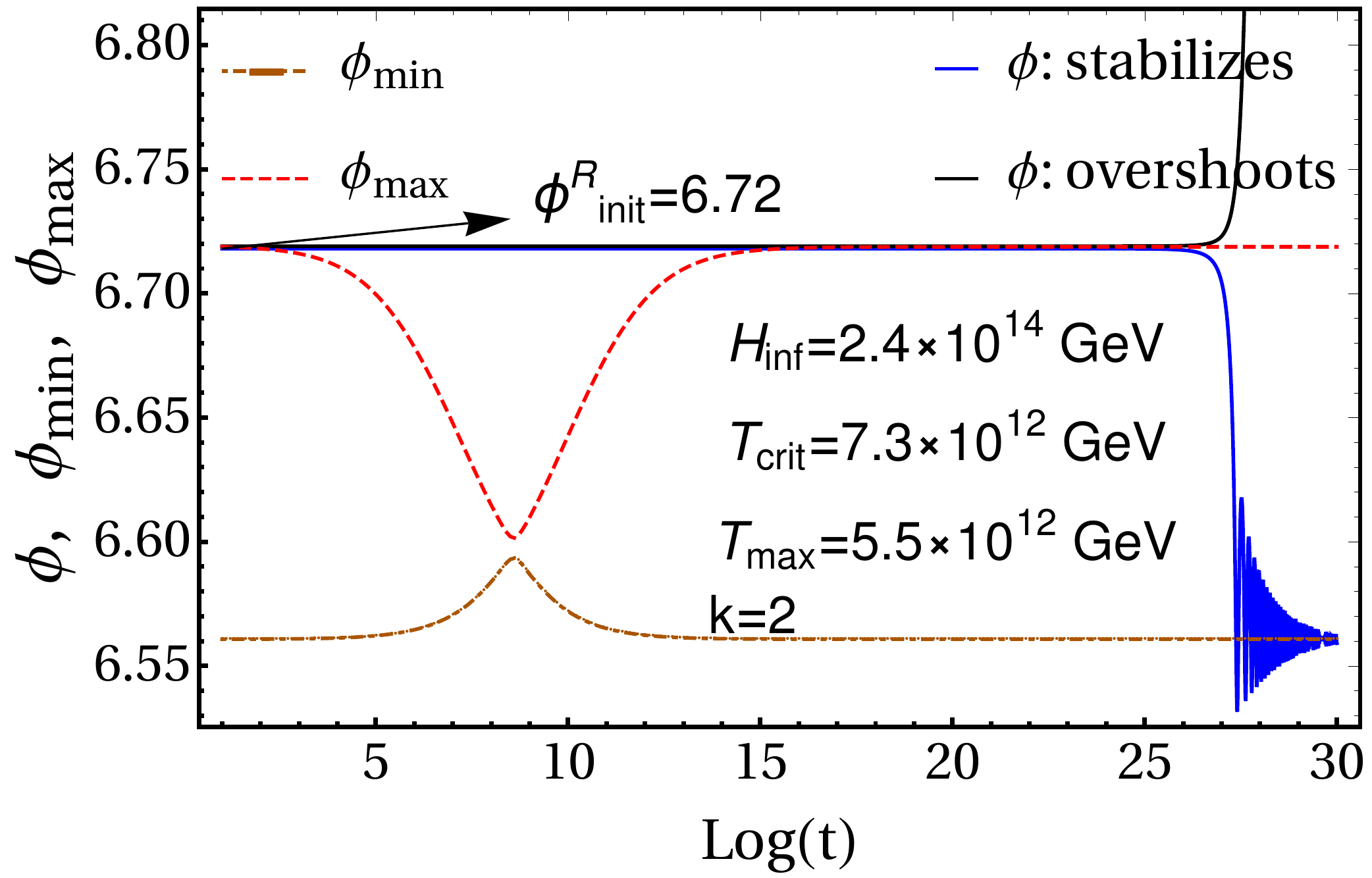}
\caption{Time evolution of the canonical modulus field during reheating is shown where the blue line corresponds to the specific initial field value for which the field does not overshoot, and the black one is for the case when the field just overshoots. The dashed red line shows the instantaneous position of the maximum of the potential, and the brown dot-dashed line corresponds to the instantaneous minimum of the potential. Both the plots are for maximum temperature ($T_{\rm max}$) smaller than the critical temperature($T_{\rm crit}$). The left panel is for $\phi_{\rm init} < \phi_{\rm min}$ and the right panel is for $\phi_{\rm init} > \phi_{\rm min}$. In the plots, we mark critical initial field values for overshooting with arrow signs.}
\label{dynamics5}
\end{figure}

To understand the effects of temperature generation via reheating, we first show the dynamics for a fixed value of $T_{\rm max}$ and contrast that with the case when the same value of the temperature was taken as an initial condition in the last section. As an example, in Fig~\ref{dynamics5}, we show the dynamics of the field for $T_{\rm max} = 5.5\times 10^{12}$ GeV. For this case, we have taken the maximum value of $H_{\rm inf}$ that is consistent with the upper limit of the tensor-to-scalar ratio produced during inflation \cite{Planck:2018jri}. In this case, $T_{\rm max}$ is just below the $T_{\rm crit} = 7.3\times 10^{12}$ GeV. From the plot, it is clear how the positions of the maximum and the minimum approach each other as the temperature rises close to the $T_{\rm crit}$ and again settle to their zero-temperature values once the radiation energy density red-shifts away. This plot should be directly contrasted with Fig.~\ref{dynamics1} that has the same initial temperature $T_{\rm init} = 5.5\times 10^{12}$ GeV. Comparing the two plots, we see that the value of $\phi^{\rm R}_{\rm init}$ is not changed much, but the value of $\phi^{L}_{\rm init}$ is changed reasonably due to the timescale of temperature generation. We also note that the field remains stuck for some time at its position even though the field is at the runaway slope. To be specific, for the field range $4.4 \leq \phi_{\rm init} \leq6.72$ the field does not overshoot whereas this range was $ 6.49 \leq \phi_{\rm init} \leq 6.7$ when the effects of radiation bath production was ignored. The value of $\phi^{\rm R}_{\rm init}$ remains the same as this value is nearly fixed by the position of the zero temperature maximum. When $T_{\rm max}$ is well below the $T_{\rm crit}$, the positions of the maximum and the minimum do not change much over the evolution of the field, and as soon as the field crosses the local maximum, the field overshoots to large field values. The overall allowed initial range is always larger than the case when the radiation density was considered as an initial condition. 

\begin{figure}[h]
	\centering
	\hspace{-0.8cm} \includegraphics[width=.49\textwidth]{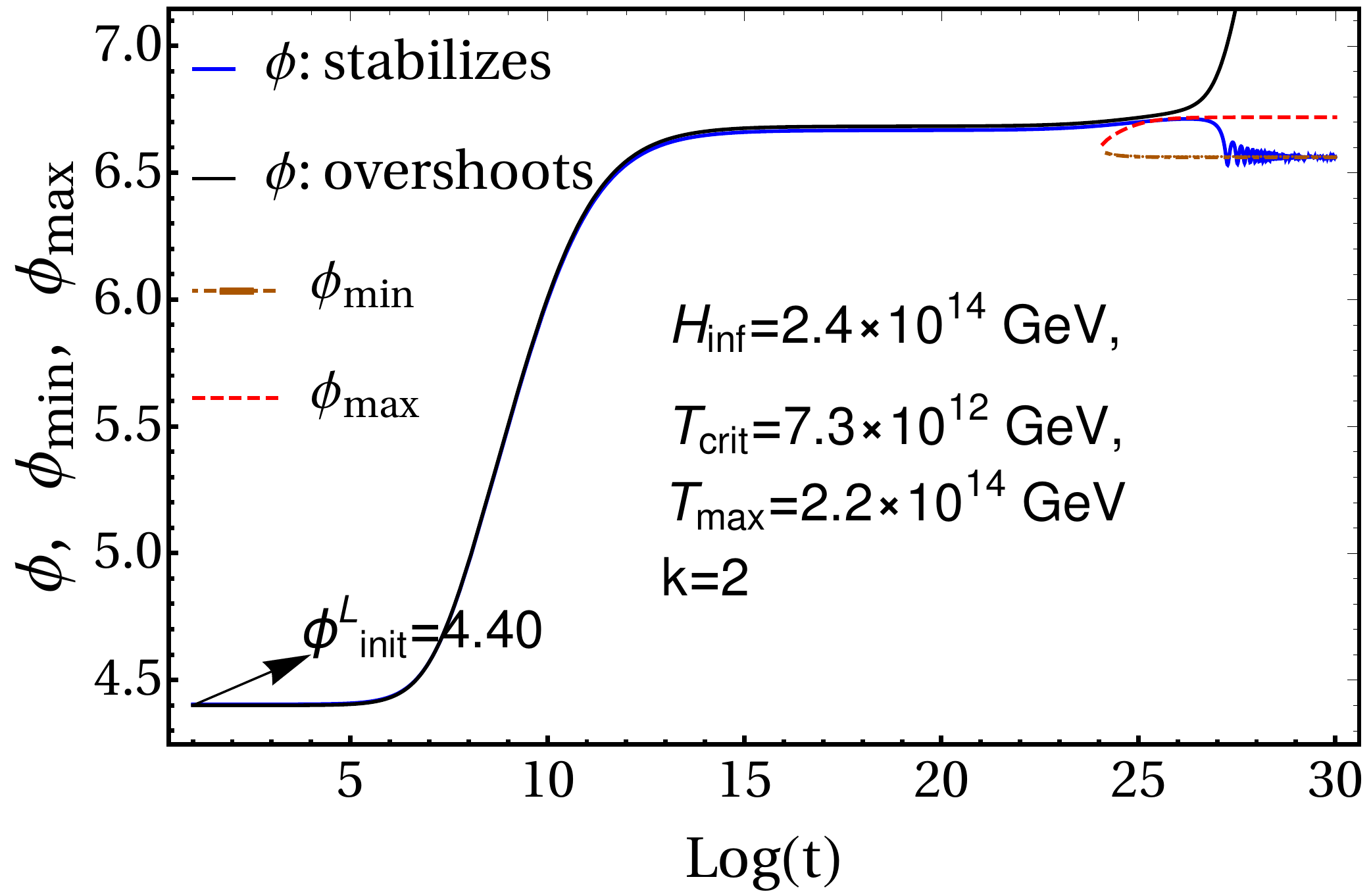}
	\hspace{.5cm}
	\includegraphics[width=0.49\textwidth]{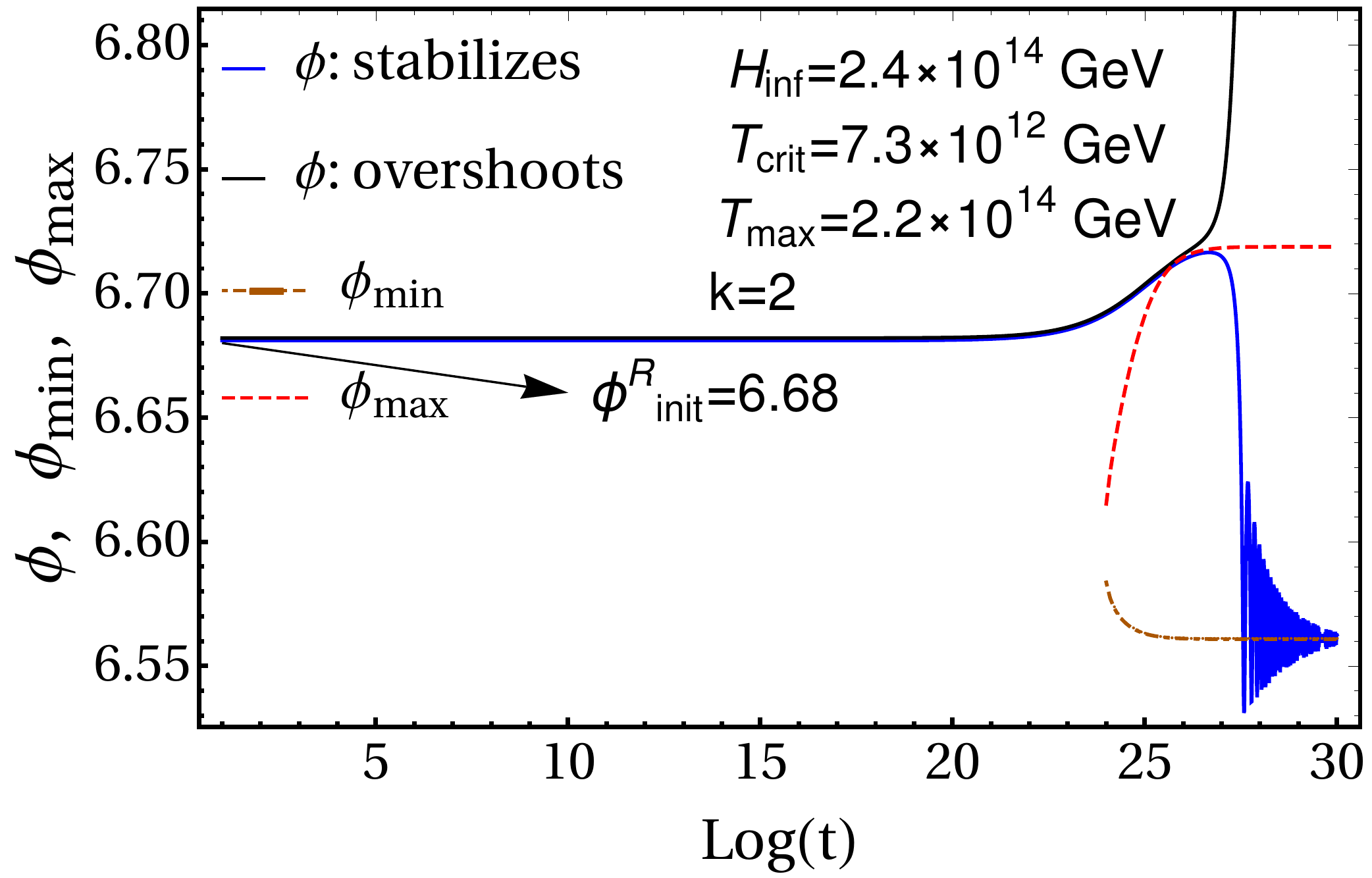}
	\caption{This is the same figure as like Fig. \ref{dynamics5}, but with maximum temperature ($T_{\rm max}$) is greater than the critical temperature ($T_{\rm crit}$).}
	\label{dynamics6}
\end{figure}

Similarly, in Fig.~\ref{dynamics6}, we show dynamics of the field when the maximum temperature produced is larger than the critical temperature. Again, in this case also the allowed initial field range is larger compared to the initial radiation bath case. The Fig.~\ref{dynamics6} should be contrasted with Fig.~\ref{dynamics3} to see the effects of continuous reheating. 

\begin{figure}[h]
\centering
\includegraphics[width=0.49\textwidth]{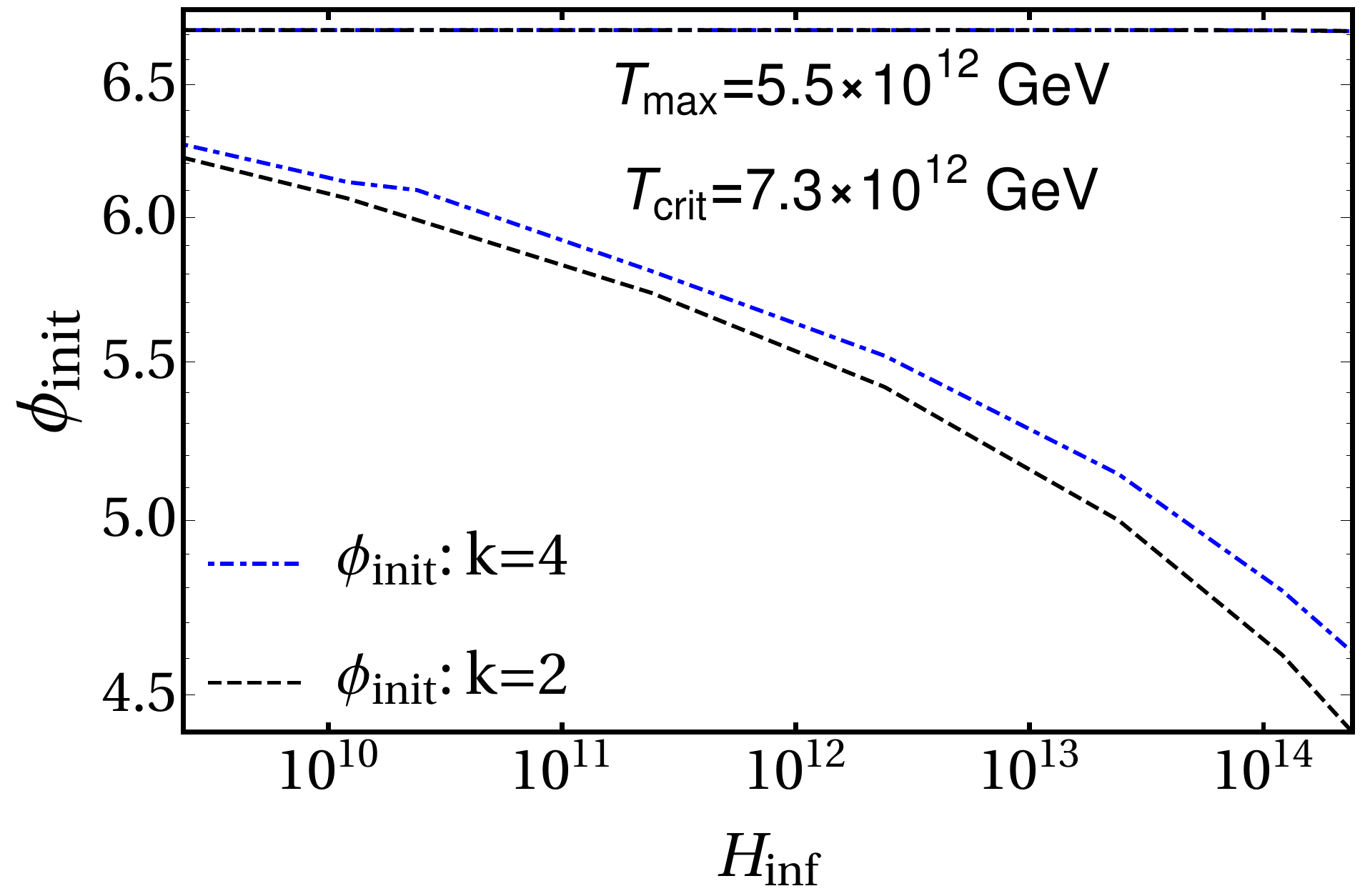}
\hspace{.5cm}
\hspace{-0.8cm} \includegraphics[width=.49\textwidth]{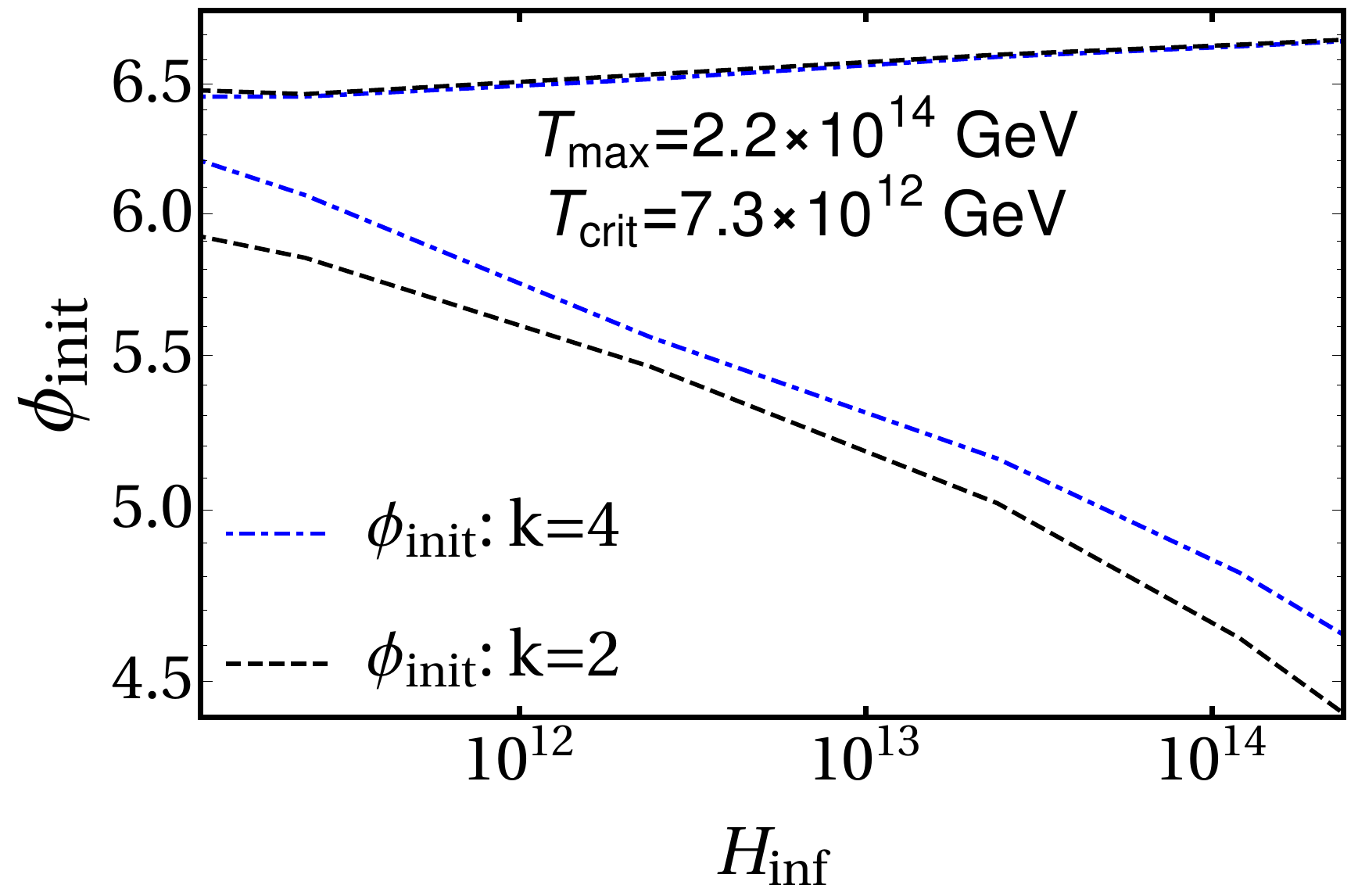}
\caption{Allowed initial field values are plotted against $H_{\rm inf}$. For the left panel $T_{\rm max}$ is smaller than the $T_{\rm crit}$, whereas for the right panel $T_{\rm max}$ is larger than $T_{\rm crit}$. The black dashed line corresponds to $k =2$, and the blue dot-dashed line is for $k = 4$. For both $k = 2$ and $k = 4$, $\phi_{\rm R}^{\rm init}$ overlaps with each other.}
\label{init_vs_H_inf}
\end{figure}

To achieve the specific value of $T_{\rm max}$, once the value of $H_{\rm inf}$ is specified, the value of $\Gamma_{\varphi}$ is also fixed. As noted above, this relaxation of the initial condition happens due to the associated time scale to produce the temperature. This is illustrated in Fig.~\ref{init_vs_H_inf}, where we show the allowed initial field values as a function of $H_{\rm inf}$ for a fixed value of $T_{\rm max}$. Note that the larger values of $H_{\rm inf}$ correspond to smaller values $\Gamma_\varphi$. For smaller values of the $\Gamma_\varphi$, the decay process of the inflaton to produce a radiation bath will take a longer time, and this will allow larger initial field space.  For all the values of $H_{\rm inf}$, the allowed field range is always larger than the case when the effects of radiation bath production are not taken into consideration. Only in the large $\Gamma_\varphi$ (i.e small $H_{\rm inf}$) limit, the radiation bath will be produced instantaneously, and we get back the results of the previous section. In  Fig.~\ref{init_vs_H_inf}, we show the field range for both $k = 2$ and $k = 4$, and we see that for $k =4$, the allowed range decreases slightly.

Till now our discussion is done for a fixed value of $r = -1.3$. To understand the effects of $r$, we show the results in Table~\ref{VAFRDVR} where $\Delta \phi^{\rm IR}$ corresponds to the case of instantaneous reheating with initial radiation bath, i.e the results of Sec.~\ref{Dynamics}, and $\Delta \phi^{\rm CR}$ corresponds to the field range where radiation is produced by continuous decay. We find no appreciable effects for the case when the initial temperature or the maximum temperature is well below the $T_{\rm crit} = 7.3\times 10^{12}$ GeV. On the other hand, when the temperature is large, the effect is slightly more prominent for larger negative values of $r$. For large negative values of $r$, the barrier height reduces, and in effect, it makes the allowed range smaller.

\begin{center}
\begin{table}
\begin{tabular}{ |p{1cm}|p{4.2cm}|p{4.2cm}|p{4.2cm}|  }
	\hline
	
	Value of r   & Allowed field range for initial radiation bath:\newline ($\Delta\phi_{\rm init}^{\rm IR}=\phi^{R}_{\rm init}-\phi^{L}_{\rm init}$) &   Allowed field range for continuous reheating: \newline ($\Delta\phi_{\rm init}^{\rm CR}=\phi^{R}_{\rm init}-\phi^{L}_{\rm init}$) & Difference between two ranges: \newline $\Delta\phi=\Delta\phi^{\rm CR}_{\rm init}-\Delta\phi_{\rm init}^{\rm IR}$\\

	\hline
	{}& $T_{\rm init}=2.2\times 10^{14} GeV$ & $T_{max}=2.2\times 10^{14} GeV, \newline H_{\rm inf}=2.4\times10^{14} GeV$ & {} \\
	\hline
	-0.7&  0.61  & 2.30 & 1.69\\
	\hline
	
	 -1.3&  0.36  & 2.27 & 1.91\\
	 \hline
	 
	 -1.7&  0.09   & 2.25 & 2.16\\
	\hline
	 
	 {}& $T_{\rm init}=3.1\times 10^{10} GeV$ & $T_{\rm max}=3.1\times 10^{10} GeV, \newline H_{\rm inf}=2.4 \times 10^{14} GeV$ & {}\\
	 \hline
	 
	 -0.7 & 0.19 & 2.33 & 2.14\\
	 \hline
	 
	 -1.3 & 0.19 & 2.33 & 2.14\\
	 \hline
	 -1.7 & 0.19 & 2.33 & 2.14\\
	 \hline
\end{tabular}
\caption{Variations of allowed field ranges for different values of $r$.}
\label{VAFRDVR}
\end{table}
\end{center}

\vspace{-0.7cm}

\begin{figure}[h]
\centering
\includegraphics[width=0.49\textwidth]{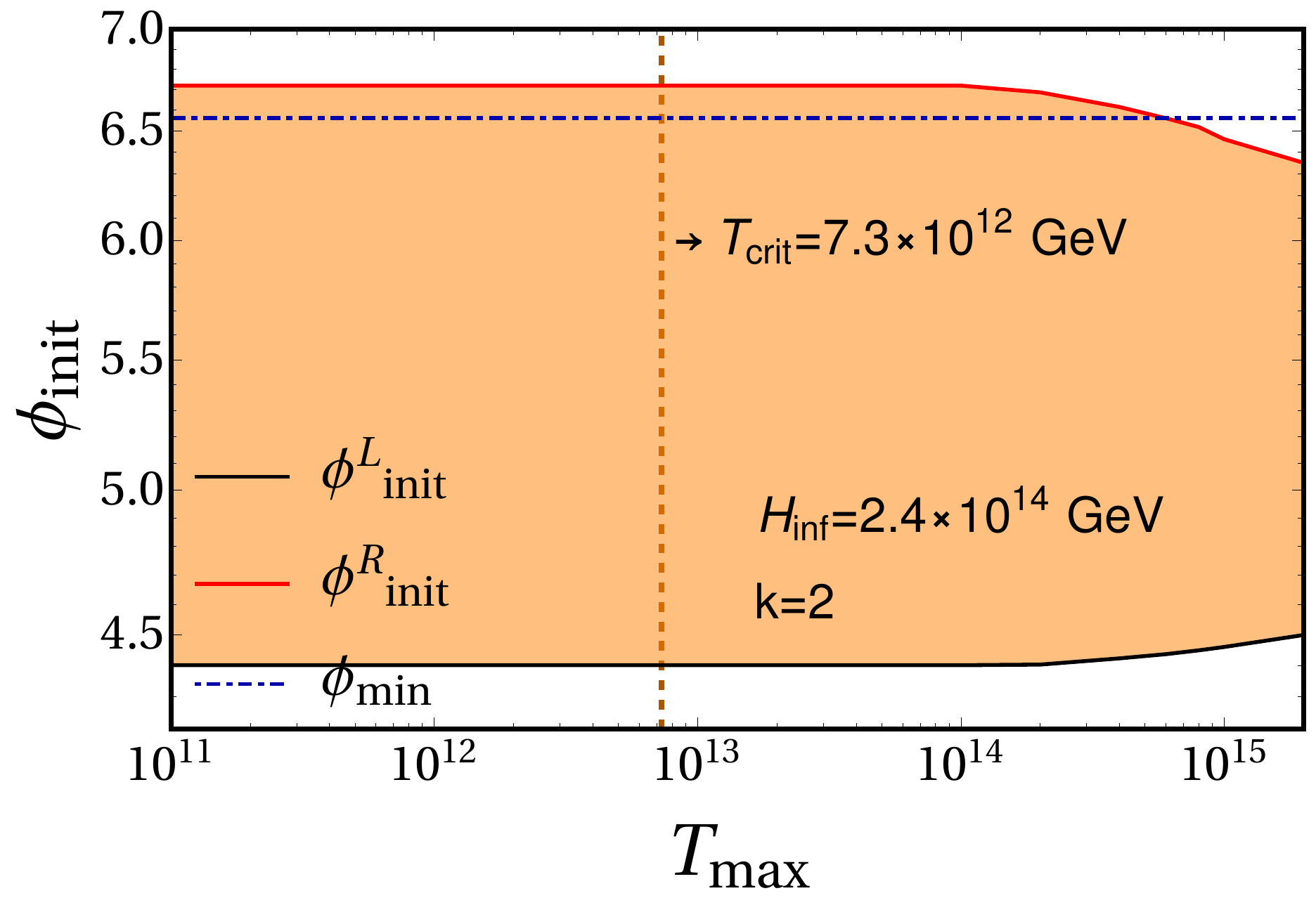}
\caption{Allowed initial field values vs $T_{\rm max}$. The vertical line corresponds to $T_{\rm crit}$, and the horizontal line marks the minimum of the zero-temperature potential.}
\label{field_Tmax}
\end{figure}

Finally, in Fig.~\ref{field_Tmax}, we show allowed initial field values when we vary $T_{\rm max}$ for a fixed value of $H_{\rm inf}$. The first thing to note is that the allowed field range does not change for a wide range of temperatures including temperatures above $T_{\rm crit}$. Therefore, not only does the modulus field not overshoot above $T_{\rm crit}$, but also the allowed field range remains roughly the same. In this range of $T_{\rm max }$, the dynamics of the field are governed by the energy density of the inflaton $ \rho_{\varphi}$. When $T_{\rm max}$ is very large, the allowed field range starts to decrease as larger $\Gamma_{\varphi}$ allows the inflaton to decay quickly. Note that for $T_{\rm max} \sim 10^{14}$ GeV (above $T_{\rm crit}$), the allowed field range remains almost the same like the lower values of $T_{\rm max}$.  In this case, the time separation between $T_{\rm max}$ and $T_{\rm R}$ is large, and the energy density is dominated by $\rho_{\varphi}$ which redshifts slower than the radiation energy density. In effect, the Hubble damping term holds for a longer time.

In summary, we conclude that the effects of reheating allow the field time to relax to its minimum without overshooting. Therefore, allowed initial field space increases compared to the case when radiation density is assumed to be present from the beginning. This effect roughly improves the allowed initial field range by one order of magnitude, see Table~\ref{VAFRDVR}, compare plots between the left panels of Fig.~\ref{dynamics4} and Fig.~\ref{field_Tmax} or Fig. \ref{dynamics1} and Fig~\ref{dynamics5}.

\section{Moduli abundance and initial conditions}
\label{moduli_abundance}

As noted in the introduction, if the mass of the moduli field $m_{\phi}$ is within the range of $10^2$-$10^3$ GeV, it decays just after the nucleosynthesis.
The most stringent bound comes from the resulting overproduction of $D~+~ _{}^{3}\textrm{He}$, and it requires that the moduli abundance relative to the entropy density $s$ at the time of reheating after the inflation should satisfy \cite{10.1143/ptp/93.5.879}, \cite{Kawasaki:2004qu}
\begin{equation}
\frac{\rho_{\phi}}{s} \lesssim 10^{-14} ~{\rm GeV} . \label{n/s_bound}
\end{equation}
If $m_{\phi} \lesssim H_{\rm inf}$, the moduli is not expected to sit at its zero temperature minimum, but it is in general shifted to a large field value $\phi_{\rm init}$ during inflation. The field begins to oscillate around its zero temperature minimum when the Hubble parameter $H$ becomes close to $m_{\phi}$. The moduli energy density $\rho_{\phi}$ divided by the entropy density $s$ is estimated as \cite{Hagihara_2019},
\begin{equation}\label{abs}
\frac{\rho_{\phi}}{s}=
\begin{cases}
\frac{1}{8}T_{R}\left(\frac{\phi_{\rm init}}{M_{p}}\right)^2& \text{for } t_{\rm osc} < t_{R}\\
\frac{1}{8}T_{\rm osc}\left(\frac{\phi_{\rm init}}{M_{p}}\right)^2              & \text{for } t_{\rm osc} > t_{R}
\end{cases}
\end{equation}
where $t_{R} (T_{R})$ is the time (temperature) at the end of reheating and $t_{osc} (T_{osc})$ is the time (temperature) at the beginning of the moduli oscillation. Here, it has been assumed that the equation of the state of the universe behaves as a non-relativistic matter before the completion of reheating produced by the inflaton. To satisfy the bound of Eq.~\eqref{n/s_bound}, the initial field value of the moduli, when it starts to oscillate, should have an upper bound \cite{Linde_1996} 
\begin{equation}\label{abs}
\phi_{\rm init} \lesssim
\begin{cases}
10^{-6} M_{p}& \text{for } t_{\rm osc} < t_{R}~,\\
10^{-10} M_{p}              & \text{for } t_{\rm osc} > t_{R}~.
\end{cases}
\end{equation}
In our analysis, we have seen that $\phi_{\rm init} \lesssim {\mathcal O} (0.1 - 0.01)$ is required to avoid the overshooting problem. On the other hand, the requirement of Eq.~\eqref{abs} is much more stringent, but only applicable to the moduli masses that cause problems for BBN light elements.

Typical moduli potentials in String theory have local minimums separated from their global minimum by a finite barrier height. For several phenomenological reasons, as discussed in the introduction, the field must be stabilized at the local minimum. The overshooting has a typical time-scale, and that is much smaller than the decay time. Therefore, preventing overshooting is absolutely necessary for all relevant moduli masses. Both for the case of zero-temperature potential or thermally corrected potential, the issue of overshooting depends on the initial conditions. For a given potential, the constraint on the initial conditions relaxes further when the effects of reheating are considered. A modulus field heavier than $\mathcal{O}(100)$ TeV decays well before the BBN, and the bound of Eq.~\eqref{abs} is not applicable, and in this case, the constraints on the initial conditions due to overshooting are applicable. On the other hand, for lighter moduli masses, as long the bound of Eq.~\eqref{abs} is satisfied, the overshooting constraints are automatically satisfied.

\section{Conclusions}

\label{Discussions and Conclusion}

In this work, we have studied the issue of moduli stabilization at the end of inflation. It is well known that constructing suitable moduli stabilizing potential is not enough to ensure that the moduli is stabilized at finite vev. It is necessary to understand the cosmological evolution of the field. Moreover, the zero-temperature potential is distorted due to the presence of radiation baths produced from the inflaton energy density. In earlier work, the dynamical analysis was done where the radiation bath was assumed as an initial condition with fixed initial temperature \cite{Barreiro_2008}. In our work, we focus on how allowed initial field space changes as the initial temperature is changed. Moreover, we discuss in detail the effects of radiation bath generation from the decay of the inflaton.

In \cite{Buchmuller:2004xr},\cite{Buchmuller:2004tz}, it was first noted that large enough thermal corrections to the potential wash away the local minimum of the potential. It was assumed that the field would destabilize immediately by reaching a large vev. This immediately puts an upper limit on the reheating temperature ($T_{\rm R}$) or the maximum temperature ($T_{\rm max}$) produced during the process of reheating. As we note, this is necessarily not the case. The issue of destabilization is always dynamical, and therefore initial field value dependent. Even for the zero temperature potential, the field destabilizes for certain initial field values \cite{Brustein:1992nk}. We find that the effects of temperature-dependent corrections do not make things worse. In fact, the allowed field space increases when the temperature is larger than the critical temperature at which the potential loses its minimum - see Fig.~\ref{dynamics4}. At the same time, when the effects of temperature generation via reheating are considered, this constraint relaxes further. The creation of a radiation bath introduces a time scale that allows the modulus field to settle more easily at its minimum. Roughly, it allows one order of magnitude more field range for stabilization.

Typical moduli potentials in String Theory have a finite barrier height, and therefore the field is always prone to overshoot that barrier if the initial conditions are not suitable. Typical overshooting time is much smaller than the decay time of modulus of all relevant masses. For heavier moduli masses ($\gtrsim 30$ TeV), the field decay before BBN, and in this case, it is absolutely necessary that the initial value of the fields are in the suitable range. On the other hand, for lighter moduli masses, even though overshooting must be avoided, the constraints coming from BBN are much more stringent. This constraint can be satisfied by tracking the field to its minimum with nearly no oscillations around \cite{Linde_1996} .

The current work can be taken further in several directions. Firstly, we considered temperature generation only via perturbative decays of the inflaton. The process can be much more complicated via non-perturbative effects like preheating etc. Those effects might be incorporated systematically. But, for our consideration, the only relevant quantity is the time-scale related to the thermal bath generation, and in the current work, it is simply parameterized by $\Gamma_{\varphi}^{-1}$.

We have noted in the introduction that a large value of inflationary potential washes away the local minimum, leading to the KL bound  \cite{Kallosh:2004yh}. Again, in this case, also, the assumption is that the field runs away to the large vevs as soon as the minima are lost. The current analysis shows that this is not necessarily the case. In a more complete analysis, the moduli field evolution needs to be studied from the time of inflation, and we leave this for future work.

In summary, we conclude that the effects of radiation bath at the end of inflation do not make the moduli stabilisation issue worse. In fact, for the temperatures larger than $T_{\rm crit}$, the allowed initial field space is similar to the zero temperature potential case.  Moreover, if the decay of the inflaton is slow to produce the bath, the field gets extra time to relax further.  
So, we can relax the upper bound of the initial temperature of the Universe or the maximum reheating temperature for a certain range of the initial field values.\\\\

\noindent
{\bf {\large Acknowledgements:}} KA is supported by IISER Kolkata doctoral fellowship. KD is partially supported by the grant MT R/2019/000395 and Indo-Russian project grant DST /INT /RUS/RSF/P-21, both funded by the DST,Govt of India.  

	

\bibliographystyle{unsrt}
\bibliography{references.bib}

\end{document}